\documentclass{aastex6}
\usepackage{bbding}
\usepackage{amssymb,amsmath}
\newcommand{\MyFigA}{\ref{MyFigA}}  \newcommand{\MyFigB}{\ref{MyFigB}}  \newcommand{\MyFigC}{\ref{MyFigC}}  \newcommand{\MyFigD}{\ref{MyFigD}}  \newcommand{\MyFigE}{\ref{MyFigE}}  \newcommand{\MyFigF}{\ref{MyFigF}}  \newcommand{\MyFigG}{\ref{MyFigG}}  \newcommand{\MyTabA}{\ref{MyTabA}}
\begin{document}
\title{Spectral Lag for a Radiating Jet Shell with a High Energy Cut-off Radiation Spectrum}
\author{Shen-Shi Du\altaffilmark{1}, Da-Bin Lin\altaffilmark{1}, Rui-Jing Lu\altaffilmark{1}, Rui-Quan Li\altaffilmark{1}, Ying-Ying Gan\altaffilmark{1}, Jia Ren\altaffilmark{1}, Xiang-Gao Wang\altaffilmark{1}, En-Wei Liang\altaffilmark{1}}
\altaffiltext{1}{Guangxi Key Laboratory for Relativistic Astrophysics, School of Physical Science and Technology, Guangxi University, Nanning 530004, People's Republic of China; lindabin@gxu.edu.cn, luruijing@gxu.edu.cn, lew@gxu.edu.cn}
\begin{abstract}
Recent research shows that the spectral lag is closely related to the spectral evolution in GRBs.
In this paper,
we study the spectral lag for a radiating jet shell with a high-energy cut-off radiation spectrum.
For the jet shell with a cut-off power-law spectrum,
the spectral lag monotonically increases with the photon energy and levels off at a certain photon energy.
It is the same for the jet shell with a Band cut-off spectrum (Bandcut).
However, a turn-over from the positive lags to negative lags appears
in the high-energy range for the jet shell with Bandcut,
which is very similar to that observed in GRB~160625B.
The dependence of the spectral lags on the spectral shape/evolution
are studied in details.
In addition, the spectral lag behavior observed in GRB~160625B
is naturally reproduced based on our theoretical outcome.
 \end{abstract}
\keywords{gamma-ray bursts: general -- individual: GRB 160625B }

\section{Introduction}\label{Sec:introduction}
Gamma-ray bursts (GRBs) are the most energetic electromagnetic explosions in the Universe.
The temporal structure of prompt gamma-ray emission exhibits diverse morphologies (\citealp{Fishman_GJ-1995-Meegan_CA-ARA&A.33.415F}),
which can vary from a single smooth large pulse to an extremely complex light curve
with many erratic overlapping pulses (see fig. 1 in \citealp{Peer_A-2015-AdAst2015E.22P}).
The radiation spectrum evolves uniformly within a GRB's pulse,
which suggests that pulses are fundamental units of GRB radiation (\citealp{Lu18}).
Thus, the observed temporal and spectral behaviors of GRB's pulses
may provide an interesting clue to understand the nature of GRBs.

The spectral lag, referring to the difference of the arrival time for different energy photons,
is a common feature of pulses in GRBs \citep{Norris86,Cheng95,Band97,Norris00,Chen05,Peng07}.
Early in the BATSE era, it was found that most of the GRBs pulses
are dominated by the positive lags (i.e., the soft photons lag behind the hard photons),
and a small fraction of pulses show negative lags (e.g., \citealt{Yi06}).
Generally, the bursts with long-wide pulses tend to have long lags and soft hardness \citep{Norris00,Norris01b,Norris05,Daigne03}.
The extensive analyses based on the GRBs observations by BATSE revealed that,
the lag features between the bursts divided by 2s-$T_{90}$-duration time \citep{Kouveliotou93}
show distinct discrepancies \citep{Band97,Norris00,Norris01a,Yi06},
which has been suggested as a new classification scheme for GRBs \citep{Norris06,Gehrels06,McBreen08,Zhang09}.
Besides, an inverse correlation between lag and the peak luminosity
was found by \cite{Norris00} based on six redshift-known bursts,
which is further confirmed by GRBs observed by BAT onboard the \emph{Swift} satellite \citep{Ukwatta10,Ukwatta12}.
This correlation is proposed to be  a GRBs distance indicator to probe cosmology \citep{Norris04,Schaefer03,Schaefer07,Gao12}.

Despite of decades of research, the true physical origin of the spectral lag is still inconclusive.
It was suggested that the curvature effect of the spherical fireball
is a plausible explanation for the spectral lag
(e.g., \citealt{Ioka01,Shen05,Lu06,Shenoy13}).
In this scenario, the emission from the spherical shells at progressively higher latitudes
with respect to the observer's line of sight
is progressively delayed due to the weaker Doppler-boosting effect,
such that the light curves of low energy radiation would be delayed and peak at later times.
But the main difficulty of curvature effect model is that the flux levels
at different energy bands are particularly lower than the observed (\citealp{Zhang09}).
\cite{Uhm16b} pointed out that there would be essentially unnoticeable
spectral lags given rise from the high-latitude emission,
as well as the properties of light curves are not in accord with the observations.
Instead, they proposed a physical model invoking a spherical shell
rapidly expands in the bulk-accelerating region, which is suggested to be more reasonable
to account for the spectral lags \citep{Uhm16b,Wei17,Shao17,Lu18,Uhm18}.
In the framework of the internal shock model, \cite{Bosnjak14} have made the thorough study of spectral evolution and investigated the spectral lags.

Recently, \cite{Lu18} reveals that the spectral lags are
strongly related to the spectral evolution in GRB's pulses,
stimulating us the motivation to explore the nature of spectral lag.
The previous works mainly focus on a radiating jet shell with a Band radiation spectrum.
However, there are a sample of GRBs of which the radiation spectrum deviates from the Band spectrum, e.g.,
cut-off power-law (e.g., GRB~170817A, \citealp{Zhang_BB-2018-Zhang_B-NatCo.9.447Z})
or Band cut-off radiation spectrum (e.g., GRB~160625B, \citealp{Lin2018}).
Thus, we would like to study the spectral lag behavior in the case that
the jet shell radiates with a high-energy cut-off spectrum based on the theory in \cite{Uhm16b}.
The contents of our paper are arranged as follows.
In Section~\ref{Sec:Sec2}, we describe the details of the physical model constructed in \cite{Uhm16b}
and explore the spectral lag behavior for GRBs with a cut-off radiation spectrum.
We also employ a phenomenological model to explore the dependence of
the spectral lag behavior on the spectral shape/evolution.
In Section~\ref{Sec:Sec3}, we discuss the spectral lags in GRB~160625B.
Our conclusions are presented in Section~\ref{Sec:Sec4}.
The flat $\Lambda$CDM cosmology with $\Omega_{\rm M}=0.27$ and $H_0=71$ km/s/Mpc is  adopted throughout this work.

\section{Spectral Lag for a Radiating Jet Shell with a Cut-off Radiation Spectrum}\label{Sec:Sec2}
\subsection{Physical Model}\label{Sec:Physical Model}
To investigate the spectral lag of GRBs' pulses,
we adopt the physical model constructed in \cite{Uhm16b}.
The physical model involves a relativistic jet shell,
which undergoes a rapid bulk acceleration and continuously emits photons
with an isotropic distribution in its co-moving frame.
Same as \cite{Uhm16b},
the radiation of the jet shell in our model
is turned on at the radius $r_{\rm{on}}=10^{14}$~cm
and turned off at $r_{\rm off}=3\times10^{16}$~cm,
where the line-of-sight emission finally ceases.
The value of $r_{\rm{on}}$ and $r_{\rm off}$ are fixed in our numerical calculations.
The synchrotron radiation is taken as the main emission mechanism
and the shape of the radiation spectrum in the shell co-moving frame is described in the form of \citep{Uhm15}
\begin{equation}
P'(E)=nP'_eH(x)\;\, {\rm with} \;\, {x=E'/{E'_{\rm{ch}}}}
\end{equation}
where $n$ represents the number density of radiating electrons,
$H$ is the normalized photon spectrum,
and $P'_e$ ($E'_{\rm{ch}}$) is the radiation spectral power (characteristic photon energy) of an emitting electron with Lorentz factor $\gamma_{\rm{ch}}$, i.e.,
\begin{equation}
P'_e= \frac{3\sqrt{3}}{32}\frac{m_e c^2\sigma_TB}{q_e},
\end{equation}
\begin{equation}
E'_{\rm{ch}}=h\nu'_{\rm{ch}}=\frac{3}{16}\frac{hq_eB}{m_ec}\gamma_{\rm{ch}}^2.
\end{equation}
Here, $m_{\rm e}$ ($q_{\rm e}$) is the mass (charge) of electrons, $c$ is the speed of light,
$\sigma_{\rm T}$ represents the Thompson cross section, $h$ is the Planck constant,
and $B$ is the strength of magnetic field in the co-moving frame.
The relativistic electrons are speculated to be uniformly distributed in the shell co-moving frame
and isotropically collected into the fluid with a constant injection rate $R_{\rm inj}=dn/dt'$.
Observationally, the sub-MeV photon spectrum of a typical GRBs can be well fitted by a smoothly joint broken power-law function (Band function; \citealt{Band93}), with a cut-off rarely observed in the high-energy bands.
In this work, we study the spectral lag behavior in the following three spectral shapes:\\
(I) the cut-off power-law radiation spectrum (CPL)
\begin{equation}
H(x)=x^{\alpha}e^{-x},
%H(E')=A\left(\frac{E'}{100\rm~keV}\right)^{\alpha}{\rm exp}\left(-\frac{E'}{E'_0}\right),
\end{equation}
(II) the Band radiation spectrum (Band)
\begin{equation}
H(x)=\left\{\begin{array}{lcl}x^{\alpha}e^{-x},&& x<(\alpha - \beta ),\\
(\alpha-\beta)^{\alpha-\beta}e^{\beta-\alpha}x^{\beta},&& x \geq (\alpha - \beta ),\\ \end{array}\right.
%H(E')=A \left\{\begin{array}{lcl}\left(\frac{E'}{100\rm~keV}\right)^{\alpha}{\rm exp}\left(-\frac{E'}{E'_0}\right),&& E'<(\alpha - \beta ) E'_0,\\
%\left[\frac{(\alpha-\beta)E'_0}{100\rm~keV}\right]^{\alpha-\beta}{\rm exp}(\beta-\alpha){\left(\frac{E'}{100\rm~keV}\right)}^{\beta},&& E'\geq (\alpha - \beta ) E'_0,\\ \end{array}\right.
\end{equation}
(III) the Band radiation spectrum with a high-energy cut-off (Bandcut)
\begin{equation}
H(x)=H^{\rm Band}(x){\rm exp}{\left(\frac{-xE'_{\rm ch}}{E'_{\rm c}}\right)},
\end{equation}
where $\alpha$ and $\beta$ are the spectral indices.
$E'_{\rm c} (\gg E'_{\rm ch})$ represents the high-energy cut-off behavior,
which may be caused by the absorption of two-photon pair production
(e.g., \citealp{Krolik_JH-1991-Pier_EA, Fenimore_EE-1993-Epstein_RI, Woods_E-1995-Loeb_A,Baring_MG-1997-Harding_AK,Ackermann_M-2011-Ajello_M, Ackermann_M-2013-Ajello_M,Tang_QW-2015-Peng_FK}).
For the above three kinds of radiation spectrum,
the peak energy of $\nu F_{\nu}$ spectrum is $E_{\rm p}=(2+\alpha)E_{\rm ch}$.
It is worthy to point out that $E_{\rm p}=(2+\alpha)E_{\rm ch}$
is only applicable for Bandcut with $E'_{\rm c}\gg E'_{\rm ch}$,
which is the cases studied in this paper.

The photons emitted in the shell co-moving frame would be Doppler-boosted with a factor of $D = 1/\left [ \Gamma \left(1-\beta \cos\theta\right)\right ]$,
where $\Gamma$ is the bulk Lorentz factor of the jet shell,
$\beta=\sqrt{1-1/\Gamma^2}$, and $\theta$ is the latitude of the emission location
relative to the observer's line of sight.
Similar to \cite{Uhm16a}, the evolution of $\Gamma$ and $B$ are described as
\begin{equation}\label{Eq:Gamma_Evolution}
\Gamma (r) = \left\{ {\begin{array}{*{20}{c}}
{{\Gamma _0},}&{{r_{{\rm{on}}}} < r \le {r_0},}\\
{{\Gamma _0}{{(r/{r_0})}^s},}&{r > {r_0},}
\end{array}} \right.
\end{equation}
\begin{equation}\label{Eq:B_Decay}
B(r) = \left\{ {\begin{array}{*{20}{c}}
{{B_0},}&{{r_{{\rm{on}}}} < r \le {r_0},}\\
{{B_0}{{(r/{r_0})}^{-b}},}&{r > {r_0},}
\end{array}} \right.
\end{equation}
where $\Gamma_0$ ($B_0$) is the normalization value at the radius $r_0$ of $\Gamma$ ($B$)
and the power-law index $s>0$ ($b>0$) is adopted \citep{Uhm14,Uhm16b}.
In this model, the jet is Poynting-flux-dominated
and the magnetic field is dissipated via reconnection of oppositely oriented field lines
as the jet propagates outward.
This is different from that in the particle-in-cell (PIC) simulations of internal shock model, in which
the simulations of shocks indicate that the Weibel instability-induced filaments merge and cause the
magnetic field to gradually decay (e.g., \citealp{Chang08,Keshet09,Silva03,Medvedev05}).
The PIC simulation of \cite{Chang08} indicates that the magnetic field decays as a power law of time
in the downstream co-moving frame (see also, e.g., \citealp{Lemoine13,Zhao14}), and the longer PIC simulation by \cite{Keshet09}
seems to result in an exponential decay with time.
The observed time $t_{\rm{obs}}$ of a photon emitted from the radius $r$ and latitude $\theta$ can be described as
\begin{equation}
t_{\rm{obs}}(r, \theta)=\left[\int_{{r_{{\rm{on}}}}}^r {\frac{{1 - \beta }}{{\beta c}}dl}+\frac{r(1-\rm cos \theta)}{c}\right](1+z).
\end{equation}
The observed flux density $F_{\nu,\rm{obs}}$ is calculated with
\begin{equation}
F_{\nu,\rm{obs}}=\int_{\rm (EATS)}{ \frac{n P'_{\rm e}D^3 H(x)(1+z)}{4\pi D^2_{\rm L}(z)}d\Omega}
\end{equation}
where EATS is the equal-arrival time surface
corresponding to the same observer time $t_{\rm obs}$
and $D_{\rm L}$ is the luminosity distance at the redshift $z$.

The spectral lags are estimated based on the discrepancy of light curves' peak time.
In this section, we adopt the following model parameters in our numerical calculations (\citealp{Uhm16b}):
$\alpha=-0.8$, $\beta=-2.3$, $\gamma_{\rm ch}=8\times10^4$, $\Gamma_0=300$, $B_0=30$~G,
$r_0=10^{15}$~cm, $s=0.35$, $b=1.25$, $R_{\rm{inj}}=10^{47}~{\rm s}^{-1}$,
and $z=1.406$ (GRB 160625B; \citealt{Xu16}).
With above parameters, the observed characteristic photon energy
in the spectrum is $E_{\rm ch}\simeq 653$~keV
at $t_{\rm obs}=0$~s and the shell curvature effect
shaping the light curves
occurs at $t_{\rm obs}\gtrsim3$~s.
For the high-energy cut-off behavior in the case with Bandcut,
we assume the cut-off energy in the co-moving frame is $E'_{\rm c}=20{\rm MeV}/2\Gamma_{\rm off}$ ( $\Gamma_{\rm off}=\Gamma_0r_{\rm off}^s/r_0^s\sim 990$)
and remains constant during the expansion of the jet shell.
Then, the observed high-energy cut-off $E_{\rm c}$ is $\simeq 6$~MeV at $t_{\rm obs}=0$~s
and increases in the case with $s>0$,
which is studied in this section.

%%%%%%%%%%%%%%%%%%%
%%%%%%%%%%%%%%%%%%%20190131
\subsection{High-energy Spectral Lag in the Physical Model}\label{Sec:numerical results}
We numerically calculate the light curves for the physical model in Section~\ref{Sec:Physical Model}.
The results are shown in Figure~{\MyFigA},
where the light curves are normalized with its peak flux $F_{\nu, \rm max}$
and the red, green, blue, purple, and black lines
are corresponding to the observed photon energy of $E_{\rm obs}=$10~keV, 100~keV, 1~MeV, 30~MeV, and 50~MeV, respectively.
The cases with CPL, Band, and Bandcut are studied in the left, middle, and right panels, respectively.
In each panel, the upper part plots the light curves at different $E_{\rm obs}$
and the lower part shows the evolution of
$E_{\rm p}$ (black dashed line) and $E_{\rm c}$ (red dashed line) if it exists.
As one can find from Figure~{\MyFigA},
the spectral lags are distinctly visible for the hard photons relative to the soft photons (e.g., 10~keV).
Moreover, the spectral lags for photons above $E_{\rm c}$ are dramatically different
for the case with Band and that with Bandcut
by comparing the purple ($E_{\rm obs}=30$~MeV) or black ($E_{\rm obs}=50$~MeV) lines.
Here, both the purple and black lines overlap the blue line ($E_{\rm obs}=1$~MeV) in the Band case
but lag behind the blue line in the Bandcut case.
Then, we study the relation of peak time $t_{\rm p}$ and $E_{\rm obs}$ in the left panel of Figure~{\MyFigB},
where the cases with CPL, Band, and Bandcut are shown with
``$\circ$'', ``{\small $\Box$}'', and ``$\times$'' symbols, respectively.
In this panel, we also plot the observed relation $t_{\rm p}\propto E_{\rm obs}^{-0.25}$ (\citealp{Band97,Liang06}) with dashed line for a comparison.
One can find that our $t_{\rm p}-E_{\rm obs}$ relations at higher
$E_{\rm obs}$ deviate from the relation of $t_{\rm p}\propto E_{\rm obs}^{-0.25}$.
For the0 cases with CPL and Band,
the $t_{\rm p}$ levels off for high $E_{\rm obs}$.
For the Bandcut case,
$t_{\rm p}$ also levels off at $E_{\rm obs}\sim 1$~MeV
but begins to rise at $E_{\rm obs}\sim 10$~MeV.
In the right panel of Figure~{\MyFigB},
we show the spectral lags $\tau$ of high-energy photons
with respect to the low-energy photons (e.g., $E_{\rm obs}=10$~keV),
i.e.,
\begin{equation}\label{Eq:tau_p}
\tau=t_{\rm p}(E_{\rm obs}=10\,{\rm keV})-t_{\rm p}(E_{\rm obs})
\footnote{In practice, the spectral lags $\tau$ are generally estimated
by using the cross-correlation function (CCF) method (see Section~\ref{Sec:Sec3}).
However, the value of $\tau$ based on CCF depends on both the peak time and pulse profile.
In order to present the effect of high-energy cut-off radiation behavior on the $\tau$,
we adopt Equation~(\ref{Eq:tau_p}) since $t_{\rm p}$ is independent on the profile of model light curves.
}.
\end{equation}
Here, the cases with CPL, Band, and Bandcut
are also shown with ``$\circ$'', ``{\small $\Box$}'', and ``$\times$'' symbols, respectively.
It can be found that the spectral lag $\tau$ increases with $E_{\rm obs}$ and
levels off at $E_{\rm obs}\gtrsim E_{\rm s}\sim 0.8\,\rm MeV$ for all cases.
However, a turn-over at $E_{\rm obs}\gtrsim E_{\rm t}\sim 10$~MeV
appears in the case with Bandcut
but does not present in the cases with CPL or Band.
This is the main finding in this work
and has been observed in GRB~160625B for the first time (\citealp{Wei17}).
We will further study this behavior in Section~\ref{Sec:physical origin}.

\subsection{Detail Study on the High-energy Spectral Lags}\label{Sec:physical origin}
\cite{Lu18} systematically studies the relation between the spectral lags and spectral evolution
based on a sample of \emph{Fermi} GRB pulses.
It is shown that the spectral lags are closely related to the spectral evolution.
In Section~\ref{Sec:numerical results},
a pattern of positive lags is found.
In addition, a high-energy turn-over in the $t_{\rm p}-E_{\rm obs}$ and $\tau-E_{\rm obs}$ relations
is presented in the case with Bandcut,
which is not studied in previous works.
Then, we would like to present a detailed study
on the relationship of the spectral lags and the spectral evolution in this section,
especially for the case with a cut-off radiation spectrum.
To be more generic for our studies,
we employ the phenomenological model in \cite{Lu18}
by giving the observed patterns of light curves and spectral evolution.
The reasons to adopt the phenomenological model are shown as follows.
(1) The calculations based on the physical model is too time-consuming.
(2) The phenomenological model can describe the GRB phenomena
without specifying any physical models.
Then, the results obtained based on the phenomenological model
is applicable for a number of GRB emission models, e.g.,
the internal shock model (\citealp{Rees94}),
the photosphere emission model (\citealp{Goodman86,Paczynski86,Thompson_C_-1994-MNRAS.270.480T,Meszaros_P-2000-Rees_MJ_-ApJ.530.292M}),
the internal-collision induced magnetic reconnection and turbulence (\citealp{Zhang_Yan11}; \citealp{Deng_Wei-2016-Zhang_Haocheng-ApJ.821L.12D}), and the external reverse shock model (e.g., \citealp{Shao_L-2005-Dai_ZG,Kobayashi_S-2007-Zhang_B,Fraija_N-2015,Fraija_N-2016-Lee_WH}).
In the phenomenological model, the observed flux density of a GRB pulse is modeled with
\begin{equation}
F_{\nu}(t_{\rm obs}, E_{\rm obs})=I_{\nu}(t_{\rm obs}) H( E_{\rm obs}/E_{\rm ch})/E_{\rm ch},
\end{equation}
where $I_{\nu}(t_{\rm obs})$ is the intensity identified with an empirical pulse model \citep{Kocevski03,Lu18}
\begin{equation}\label{Eq:LC}
I_{\nu}(t_{\rm obs})=I_{\rm p,f} \left(\frac{t_{\rm obs}-t_0}{t_{\rm p,f}-t_0}\right)^{\mathcal{R}}
\left[\frac{\mathcal{D}}{\mathcal{D}+\mathcal{R}}+\frac{\mathcal{R}}{\mathcal{D}+\mathcal{R}}
\left(\frac{t_{\rm obs}-t_0}{t_{\rm p,f}-t_0}\right)^{(\mathcal{R}+1)}\right]^{-\frac{\mathcal{R}+\mathcal{D}}{\mathcal{R}+1}}.
\end{equation}
Here, $t_0$ corresponds to the zero time of pulse,
and $\mathcal{R}$ and $\mathcal{D}$ are the power-law indices before
and after the time ($t_{\rm p,f}$) of the peak flux ($I_{\rm p,f}$) in full-wave band.
For the temporal evolution of $E_{\rm p}=(2+\alpha)E_{\rm ch}$, we adopt the hard-to-soft mode (\citealp{Lu18}), i.e.,
\begin{equation}\label{Eq:Ep_evolution}
E_{\rm p}(t_{\rm obs})= E_{\rm p,0}\left(1+\frac{t_{\rm obs}-t_0}{t_{\rm c}}\right)^{- k_{\rm p}}\;\text{ with }\;k_{\rm p}>0,
\end{equation}
where $t_{\rm c}$ is the characteristic timescale of the $E_{\rm p}$'s evolution.
In observations, the observed high-energy cut-off $E_{\rm c}$ may also evolve with time,
e.g., the value of $E_{\rm c}$ in the second sub-burst of GRB~160625B increases with $t_{\rm obs}$
(\citealp{Lin2018}).
Then, the temporal evolution of $E_{\rm c}$ in the Bandcut case
is set to be
\begin{equation}\label{Eq:Ec_evolution}
E_{\rm c}(t_{\rm obs})= E_{\rm c,0}\left(1+\frac{t_{\rm obs}-t_0}{t_{\rm c}}\right)^{k_{\rm c}}\; \text{ with }\;k_{\rm c}>0.
\end{equation}
In this section, we adopt the following model parameters in our numerical calculations:
$I_{\rm p,f}=1$, $\mathcal{R}=1$, $\mathcal{D}=3$, $t_0=0$~s, $t_{\rm p,f}=5$~s,
$E_{\rm p,0}=1$~MeV, $E_{\rm c,0}=20$~MeV, $k_{\rm p}=1$, and $k_{\rm c}=0.5$, respectively.
With these parameters, the rise timescale of the pulse in bolometric luminosity is around $3.5$~s and thus
the value of $t_c=3.5$~s is adopted.

With above phenomenological descriptions,
we explore the spectral lag behavior for the cases with a high-energy cut-off radiation spectrum,
i.e., CPL and Bandcut.
The synthetic light curves (upper part) as well as the spectral evolution (lower part) are shown
in the left (CPL case) and middle (Bandcut case) panels of Figure~{\MyFigC},
where the light curves observed at the photon energy $E_{\rm obs}=10$~keV,
100~keV, 1~MeV, 30~MeV, and 50~MeV are plotted with red, green, blue, purple, and black lines, respectively.
For the lower part of middle panel, the evolution of $E_{\rm p}$ ($E_{\rm c}$) is drawn with dashed black (red) line.
In addition, the peak time of each light curve is denoted by the same color vertical dashed line.
According to these light curves, one can find the very different spectral lag behaviors
of $E_{\rm obs}>E_{\rm c}$ (e.g., purple or black line)
for the case with CPL and that with Bandcut.
This result is consistent with that found in Section~\ref{Sec:numerical results}.
The right panel of Figure~{\MyFigC} displays the relations of $\tau-E_{\rm obs}$,
where ``$\circ$'' and ``$\times$'' symbols denote the results from the cases with CPL and that with Bandcut, respectively.
The relations of $\tau-E_{\rm obs}$ in this panel
are also consistent with those obtained in Section~\ref{Sec:numerical results}.
Especially, a turn-over of $\tau-E_{\rm obs}$ relation in the high-energy channels
also appears in the Bandcut case, which is what we mainly expect.
These results suggest that the spectral lag revealed by the phenomenological model
is consistent with that estimated based on the physical model.

Based on the $\tau-E_{\rm obs}$ relations in the right panels of Figures~{\MyFigB} and {\MyFigC},
we formulate the spectral lag behavior as follows,
\begin{equation}\label{Eq:tau_Formula}
\tau=
\begin{cases}
\tau_0 \frac{E_{\rm obs}^{k_{\rm s}}-E^{k_{\rm s}}_{\rm m}}{E^{k_{\rm s}}_{\rm s}-E^{k_{\rm s}}_{\rm m}}, &  E_{\rm obs}\leqslant E_{\rm s}\\
\tau_0\left[1+\left(\frac{E_{\rm obs}}{E_{\rm t}}\right)^{k_{\rm t} \omega}\right]^{\frac{1}{\omega}}, &  E_{\rm obs}> E_{\rm s}
\end{cases}
\end{equation}
where $\tau_0$ is the spectral lag at $E_{\rm obs}=E_{\rm s}$,
$E_{\rm m}$ is the specified lowest energy (e.g., 10~keV),
$k_{\rm s}$ ($k_{\rm t}$) is the power-law index for
the energy range $E_{\rm obs}<E_{\rm s}$ ($E_{\rm obs}>E_{\rm t}$),
and $\omega$ describes the sharpness around the break $E_{\rm t}$ (actually $\omega=1$ is adopted in this work).
Equation~(\ref{Eq:tau_Formula}) is applicable for the case with CPL or Band by setting $E_{\rm t}$ significantly high (e.g., $E_{\rm t}\rightarrow \infty$).
The dashed lines in right panels of Figures~{\MyFigB} and {\MyFigC}
are the fitting results based on Equation~(\ref{Eq:tau_Formula}),
where the cases with CPL, Band, and Bandcut are shown with the blue, green, and red color.
It can be found that Equation~(\ref{Eq:tau_Formula}) well describes the relations of $\tau-E_{\rm obs}$.

In this part, we investigate the dependence of the spectral lag behavior
on the radiation spectral shape/evolution for the case with CPL or Bandcut.
We first study the case with Bandcut and the results are shown in
Figure~{\MyFigD}.
Here, the upper-left (right) panel displays the relations between
the spectral lag behavior (i.e., $E_{\rm s}$, $k_{\rm s}$, $\tau_0$, $E_{\rm t}$, and $k_{\rm t}$)
and the spectral index $\alpha$ ($\beta$) by
adopting $E_{\rm p,0}=1$~MeV, $k_{\rm p}=1$, $E_{\rm c,0}=20$~MeV, and $k_{\rm c}=0.5$.
One can find that
the value of $\tau_0$ and thus the spectral lag increases with $\alpha$ while decreases with $\beta$.
In addition, the $\tau-E_{\rm obs}$ relation in the energy range of $E_{\rm obs}<E_{\rm s}$
becomes shallower (steeper) by increasing the value of $\alpha$ ($\beta$).
However, the $k_{\rm t}$ is not related to the spectral shape.
For the two characteristic photon energy in the $\tau-E_{\rm obs}$ relations,
$E_{\rm s}$ decreases by increasing $\alpha$ or $\beta$,
and $E_{\rm t}$ increases with $\alpha$ and decreases with $\beta$.
In the middle-left (right) panel of Figure~{\MyFigD},
we show the relations between
the spectral lag behavior (i.e., $E_{\rm s}$, $k_{\rm s}$, $\tau_0$, $E_{\rm t}$, and $k_{\rm t}$)
and the spectral evolution index $k_{\rm p}$ ($k_{\rm c}$)
by setting $\alpha=-0.8$ and $\beta=-2.3$.
Here, $E_{\rm p,0}=0.5$, 1, 2~MeV ($E_{\rm c,0}=5$, 10, 20~MeV) are adopted in the left (right) panel
and plotted with ``$\circ$'', ``$\diamond$'', and ``$\ast$'' symbols, respectively.
From these panels, one can find that
the values of $E_{\rm s}$, $k_{\rm s}$, and $\tau_0$
are related to the value of $k_{\rm p}$
but does not depend on the $k_{\rm c}$.
It reveals that the values of $E_{\rm s}$, $k_{\rm s}$, and $\tau_0$
are associated with the Band spectrum part of Bandcut rather
than the high-energy cut-off behavior.
Moreover, the $\tau-E_{\rm obs}$ relation in the energy range of $E_{\rm obs}<E_{\rm s}$
becomes shallower by increasing the value of $k_{\rm p}$.
The spectral lag becomes larger by adopting high value of $k_{\rm p}$,
which is consistent with the observed one (e.g., the left panel of figure 6 in \citealp{Lu18}).
The value of $k_{\rm p}$ and $k_{\rm c}$ also affects the turn-over behavior of the $\tau-E_{\rm obs}$ relations in the high-energy range.
The higher value of $k_{\rm p}$ or $k_{\rm c}$ adopted,
the higher value of $E_{\rm t}$ would be.
In addition, a low value of $k_{\rm t}$ would be produced in the case with high $k_{\rm c}$.
It is interesting to point out that
the value of $E_{\rm p,0}$ ($E_{\rm c,0}$) only influences the value of $E_{\rm s}$ ($E_{\rm t}$).
Then, we plot the relation of $E_{\rm s}-E_{\rm p,0}$ ($E_{\rm t}-E_{\rm c,0}$)
in the lower-left (right) panel.
One can find that $E_{\rm s}$ ($E_{\rm t}$) is proportional to $E_{\rm p,0}$ ($E_{\rm c,0}$), which is not associated to the $k_{\rm p}$ ($k_{\rm c}$).
These results indicate
that the spectral lag is strongly related to the spectral shape and evolution.
We also investigate the dependence of the spectral lag behavior
on the radiation spectral shape/evolution for the case with CPL.
The results are shown in Figure~{\MyFigE}.
For the cases with CPL, the $\tau-E_{\rm obs}$ relation can be described with three parameters, i.e.,
$E_{\rm s}$, $k_{\rm s}$, and $\tau_0$
{\footnote{Here, $E_{\rm t}=+\infty$ is adopted in our fitting on the $\tau-E_{\rm obs}$ relation.}}.
The left and middle panels of Figures~{\MyFigE}
show the dependence of the above three parameters on the $\alpha$
and $k_{\rm p}$, respectively.
The dependences of $E_{\rm s}$, $k_{\rm s}$, and $\tau_0$ on the value of $\alpha$ ($k_{\rm p}$)
are almost consistent with those found in the Bandcut case.
In addition, the value of $E_{\rm s}$ is also proportional to the value of $E_{\rm p,0}$.
This behavior is also consistent with that found in the Bandcut case.

%%%%%%%%%%%%%%%%%%%%%%%%%%%%%%0%%%%%%%%%%%%%%%%%%%%%%%%%%%%%%%%%%%%%%%%%%%%%%%%%%%%%%%%%%%%%%%%%%%%%%%%%%
%%%%%%%%%%%%%%%%%%%%%%%%%%%%%%%%%%%%%%%%%%%%%%%%%%%%%%%%%%%%%%%%%%%%%%%%%%%%%%%%%%%%%%%%%%%%%%%%%%%%%%%%
\section{Discussion: Application to GRB~160625B}\label{Sec:Sec3}
The prompt $\gamma$-ray emission of GRB~160625B consists of three distinct emission episodes
with a total duration of about $T_{90}=770$~s (15-350~keV; \citealt{Zhang18}).
The first sub-burst (episode I) that triggered the \emph{Fermi}/GBM at $T_0$=22:40:16.28~UT on 2016 June 25
lasts approximately 0.8~s with soft radiation spectrum.
At $T_0+188.54$~s, the \emph{Fermi}/LAT detected the main sub-burst (episode II),
which is an extremely bright episode with multiple peaks and a duration of about 35~s.
The main sub-burst is also detected by the GBM detector.
After a long quiescent stage of 339~s, the GBM was triggered again,
resulting in the third sub-burst (episode III) with a duration of about 212~s.
The spectroscopic observations of the absorptions lines are coincident
with Mg I, Mg II, and Fe II at a common redshift $z=1.406$ \citep{Xu16}.

Interestingly, the first pulse of the main sub-burst in GRB~160625B is structure-smooth
and extremely bright (see the pulse enclosed by red lines in the left panel of Figure~{\MyFigF}).
We can therefore extract the light curves in different energy channels
and calculate the spectral lags by using the cross-correlation function (CCF) method (see e.g., \citealt{Cheng95,Zhang12}).
The spectral lag is estimated with respect to the lowest energy band
($10-25$~keV) and
its uncertainties are estimated by Monte Carlo simulations (see e.g., \citealt{Ukwatta10,Zhang12,Lu18}).
The results are reported in Table~{\MyTabA} and shown in Figure~{\MyFigG} with ``$\bullet$'' symbol.
The $\tau$ increases with respect to $E_{\rm obs}$
and levels off at $E_{\rm s}\sim850$~keV.
In particular, a turn-over appears at $E_{\rm t}\simeq 40$~MeV.
This behavior is in accord with our theoretical outcome
for the case with Bandcut and a hard-to-soft spectrum evolution
(e.g., the right panels of Figures~{\MyFigB} and {\MyFigC}).
In the other hand, we note that
the time-resolved spectrum of the first pulse in the main sub-burst of GRB~160625B
can be well described with a Band cut-off radiation spectrum (\citealp{Lin2018}).
The evolution of $\alpha$, $\beta$,
$E_{\rm p}$, and $E_{\rm c}$ can be found in the right panel of Figure~{\MyFigF},
where the red solid lines indicate the fitting results.
The fitting results are also shown in each sub-figure.
As indicated in Section~\ref{Sec:physical origin},
the spectral lags strongly depend on the evolution of spectral indices, $E_{\rm p}$, and $E_{\rm c}$.
Based on the fitting results in Figure~{\MyFigF},
we numerically calculate the spectral lags based on the phenomenological model
with a Band cut-off radiation spectrum.
The result is plotted in Figure~{\MyFigG} with ``$\circ$'' symbol.
One can find that our numerical spectral lags are well consistent with the observations.
Especially, a turn-over also appears in the high-energy range.
It is suggested that the increase of the time lag
in the energy range $E_{\rm obs}\in(850{\,\rm keV},\,40{\,\rm MeV})$
is related to the evolution of spectral indices in the first pulse of GRB~160625B main sub-burst.
In addition, the turn-over in the energy range $E_{\rm obs}\gtrsim40{\,\rm keV}$
is associated with the high-energy cut-off of the Band cut-off radiation spectrum.

\section{Conclusions}\label{Sec:Sec4}
This paper focuses on the spectral lag behavior for an radiating jet with a high-energy cut-off radiation spectrum.
Based on the physical model constructed in \cite{Uhm16b},
we find that the spectral lag $\tau$ monotonically increases with photon energy $E_{\rm obs}$
and levels off at a certain $E_{\rm obs}$
in the case with CPL/Band and hard-to-soft spectral evolution.
This behavior is consistent with the previous works (e.g., \citealp{Lu06,Peng11}).
In particular, we find a turn-over from the positive lags to negative lags
in the high-energy range for the case with a Bandcut.
Such kind of the spectral lags are also reproduced based on the phenomenological model
(see Section~\ref{Sec:physical origin}).
For our obtained results, we come up with a reasonable formulation to describe
the $\tau-E_{\rm obs}$ relations.
Then, we perform further investigation on the relations between the spectral lags
and the spectral shape/evolution.
Moreover, the spectral lags observed in GRB~160625B
and the $\tau-E_{\rm obs}$ relation can be naturally reproduced
by adopting the phenomenological model with Bandcut.
Then, one can conclude that the spectral lag strongly depends on
both the spectral shape and spectral evolution in pulses of GRBs.

%%%%%%%%%%%%%%%%%%%%%%%%%%%%%%%%%%%%%%%%%%%%%%%%%%%%%%%%%%%%%%%%%%%%%%%%%%%%%%%%%%%%%%%%%%%%%%%%%%%%%%%%%
%%%%%%%%%%%%%%%%%%%%%%%%%%%%%%%%%%%%%%%%%%%%%%%%%%%%%%%%%%%%%%%%%%%%%%%%%%%%%%%%%%%%%%%%%%%%%%%%%%%%%%%%%
\acknowledgments
We thank the anonymous referee of this work for beneficial suggestions that improved the paper.
We also thank Shan-Qin Wang for beneficial discussions.
This work is supported by
the National Natural Science Foundation of China
(grant Nos. 11773007, 11533003, 11851304, U1731239),
the Guangxi Science Foundation
(grant Nos. 2018GXNSFFA281010, 2016GXNSFDA380027, 2018GXNSFDA281033, 2017AD22006, 2018GXNSFGA281005),
the Innovation Team and Outstanding Scholar Program in Guangxi Colleges,
the One-Hundred-Talents Program of Guangxi colleges,
and the Innovation Project of Guangxi Graduate Education (grant Nos. YCSW2018050).

%%%%%%%%%%%%%%%%%%%%%%%%%%%%%%%%%%%%%%%%%%%%%%%%%%%%%%%%%%%%%%%%%%%%%%%%%%%%%%%%%%%%%%%%%%%%%%%%%%%%%%%%%%%%%%%%%%%%%%%%%%%%%%%%%%%%%%%%%%%%%%%%%%%%%%%%%%
%%%%%%%%%%%%%%%%%%%%%%%%%%%%%%%%%%%%%%%%%%%%%%%%%%%%%%%%%%%%%%%0%%%%%%%%%0%%%%%%%%%%%%%%%%%%%%%%%%%%%%%%%%%%%%%%%%%%%%%%%%%%%%%%%%%%%%%%%%%%%%%%%%%%%%%%%%%%

\clearpage
\begin{figure}
\begin{center}
\begin{tabular}{ccc}
\includegraphics[angle=0,scale=0.15]{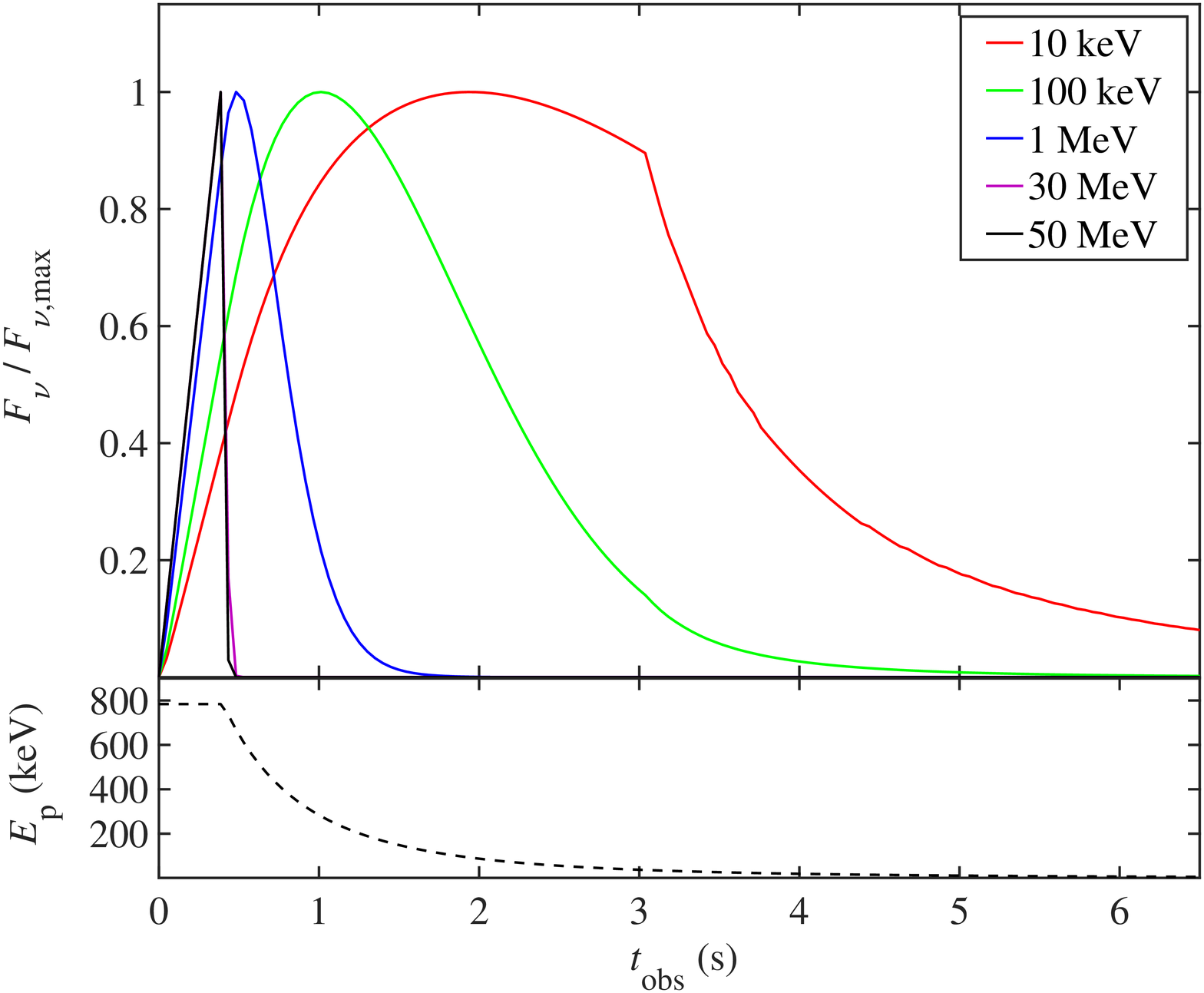} &
\includegraphics[angle=0,scale=0.15]{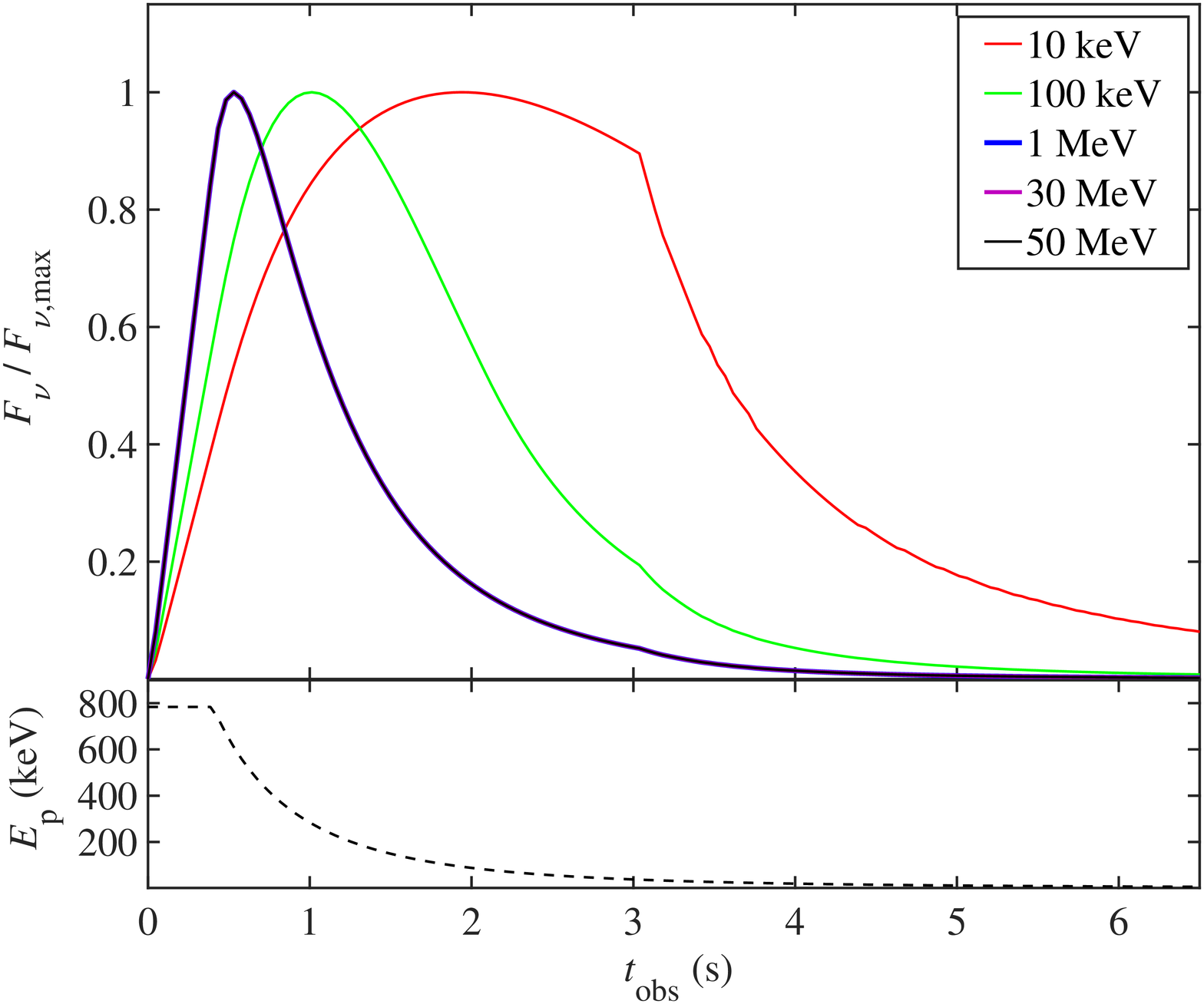} &
\includegraphics[angle=0,scale=0.15]{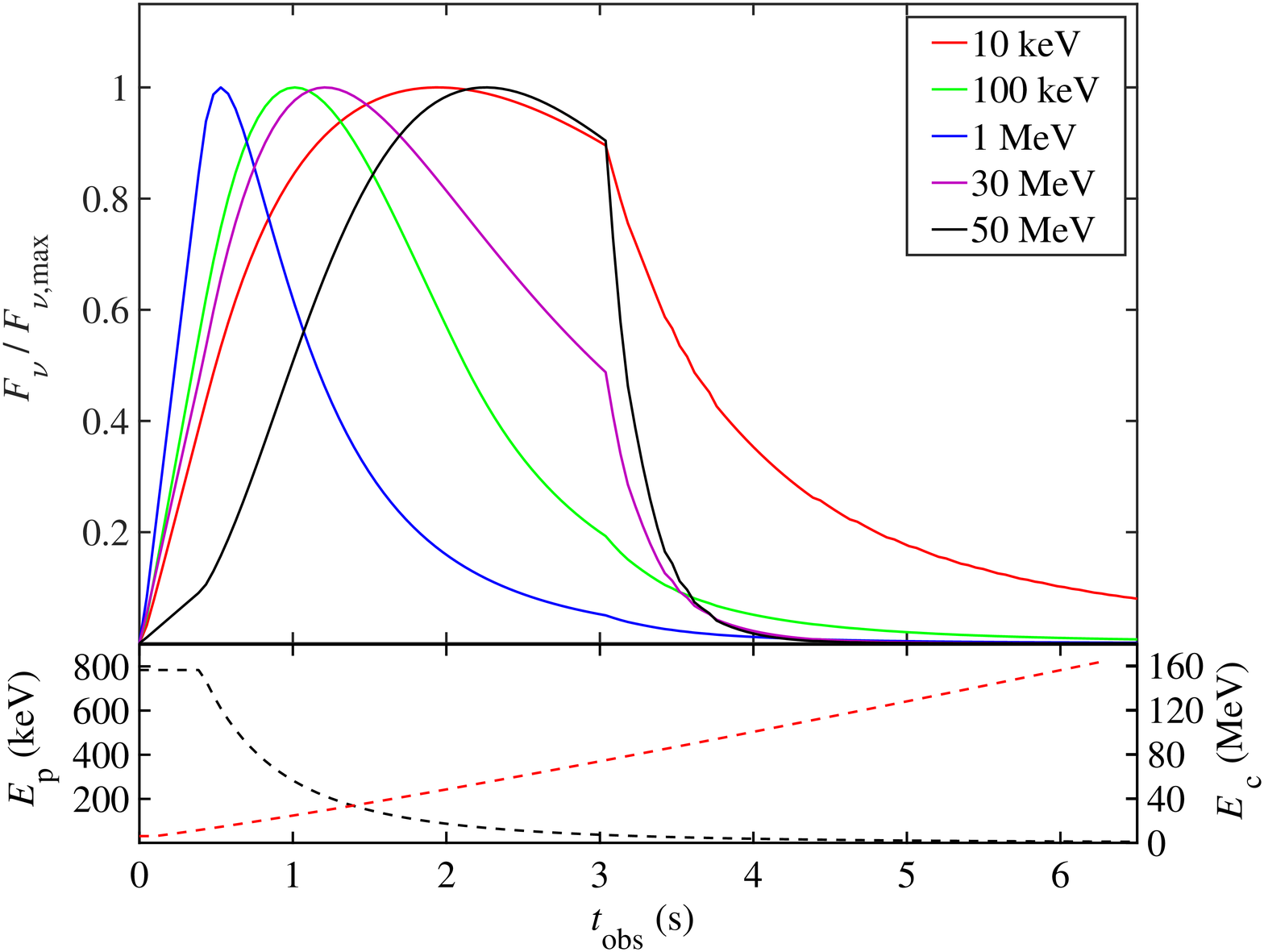}\\
\end{tabular}
\end{center}
\caption{The simulated light curves (upper sub-figure) and $E_{\rm ch}$ or $E_{\rm p}$ ($E_{\rm c}$) evolution (lower sub-figure) for the physical model (see Section~\ref{Sec:Physical Model}) with
CPL (left panel), Band (middle panel), and Bandcut (right panel) radiating photon spectrum, respectively.
The red, green, blue, purple, and, black lines
are corresponding to the light curves observed at $E_{\rm obs}=$10~keV, 100~keV, 1~MeV, 30~MeV, and 50~MeV, respectively.
}\label{MyFigA}
\end{figure}
%%%%%%%%%%%%%%%%%%%%%%%%%%%%%%%%%%%%%%%0%%%%0%%%%%%%%%0%%%%%%%%%%%%%%%%%%%%%%%%%%%%%%
%%%%%%%%%%%%%%%%%%%%%%%%%%%%%%%%%%%%%%%0%%%%0%0%%%%%%0%%%%%%%%%%%%%%%%%%%%%%%%%%%%%%%%%

\clearpage
\begin{figure}
\begin{center}
\begin{tabular}{cc}
\includegraphics[angle=0,scale=0.33]{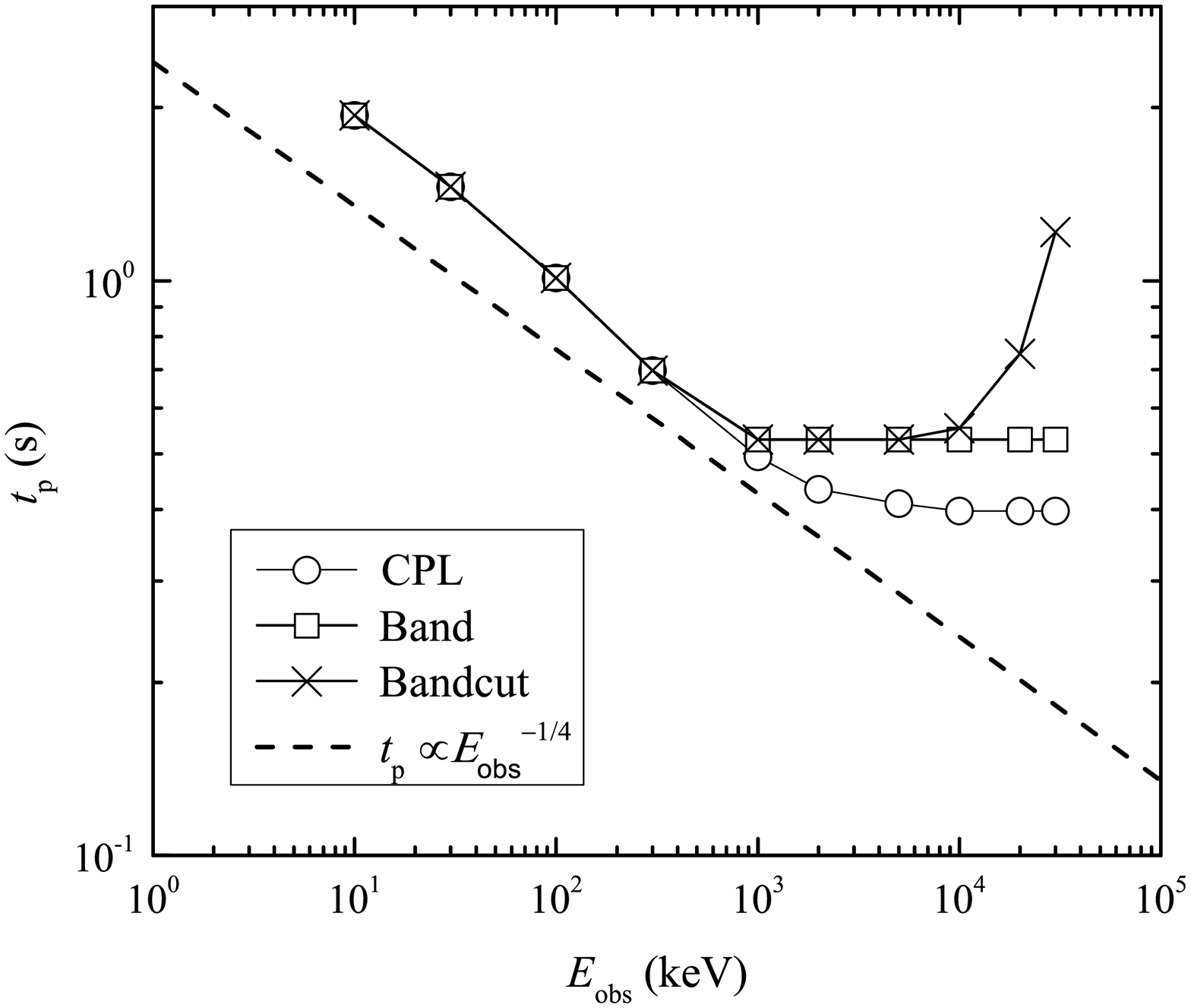} &
\includegraphics[angle=0,scale=0.33]{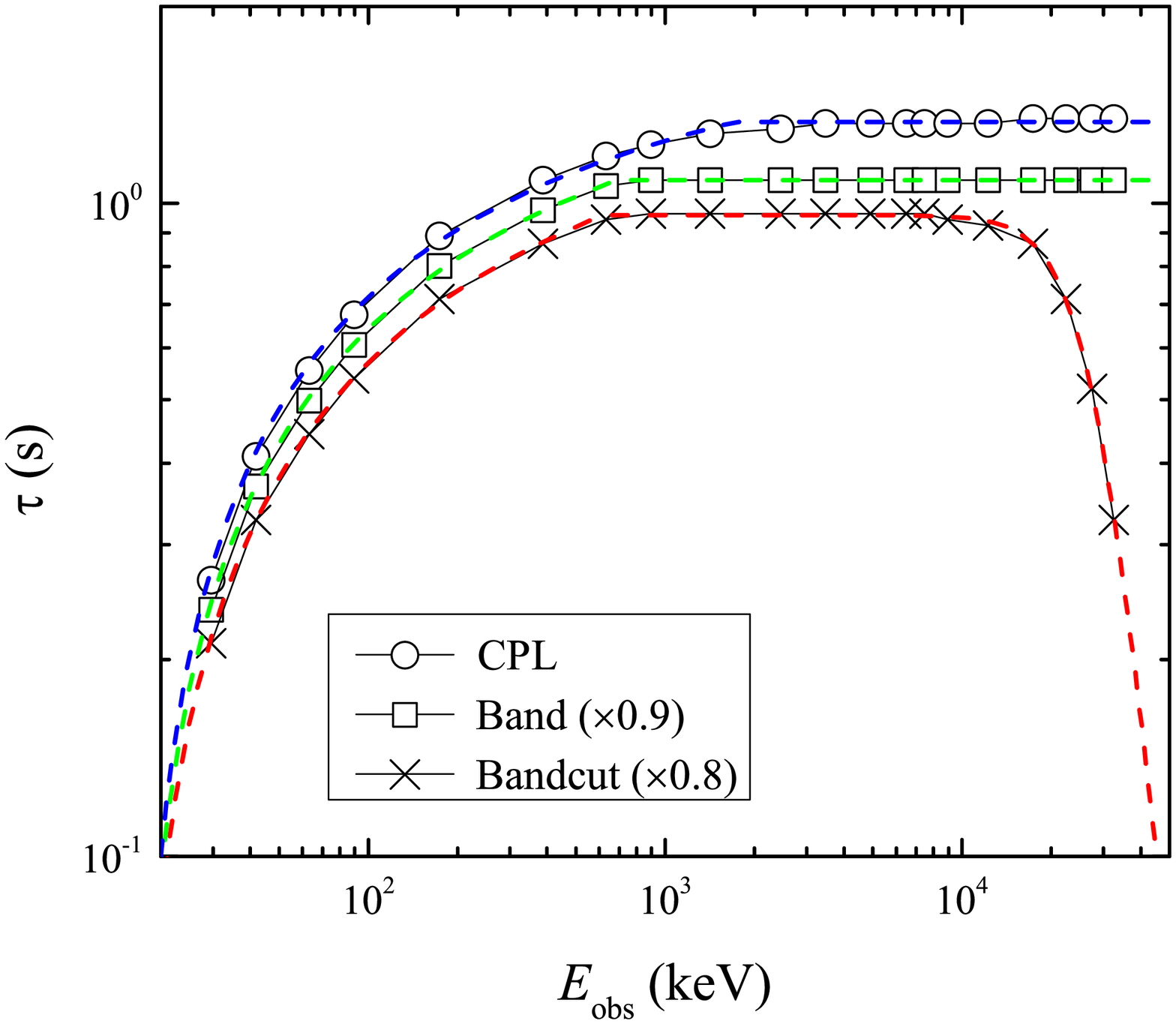}\\
\end{tabular}
\end{center}
\caption{$t_{\rm p}-E_{\rm obs}$ (left panel) and $\tau-E_{\rm obs}$ (right panel) relations,
where the situations with CPL, Band, and Bandcut radiation spectrum
are shown with the symbols $\circ$, $\square$, and $\times$, respectively.
In addition, the relation of $t_{\rm p} \propto \nu_{\rm obs}^{-0.25}$
is also plotted in the left panel (dashed line) for comparisons. The dashed lines in the right panel are well fitted by using Equation~(\ref{Eq:tau_Formula}).}
\label{MyFigB}
\end{figure}
%%%%%%%%%%%%%%%%%%%%%%%%%%%%%%%%%%%%%%%0%%%%0%%%%%%%%%0%%%%%%%%%%%%%%%%%%%%%%%%%%%%%%
%%%%%%%%%%%%%%%%%%%%%%%%%%%%%%%%%%%%%%%%%%%%0%%%%%%0%%%%%%%%%%%%%%%%%%%%%%%%%%%%%%%%%

\clearpage
\begin{figure}
\begin{center}
\begin{tabular}{ccc}
\includegraphics[angle=0,scale=0.15]{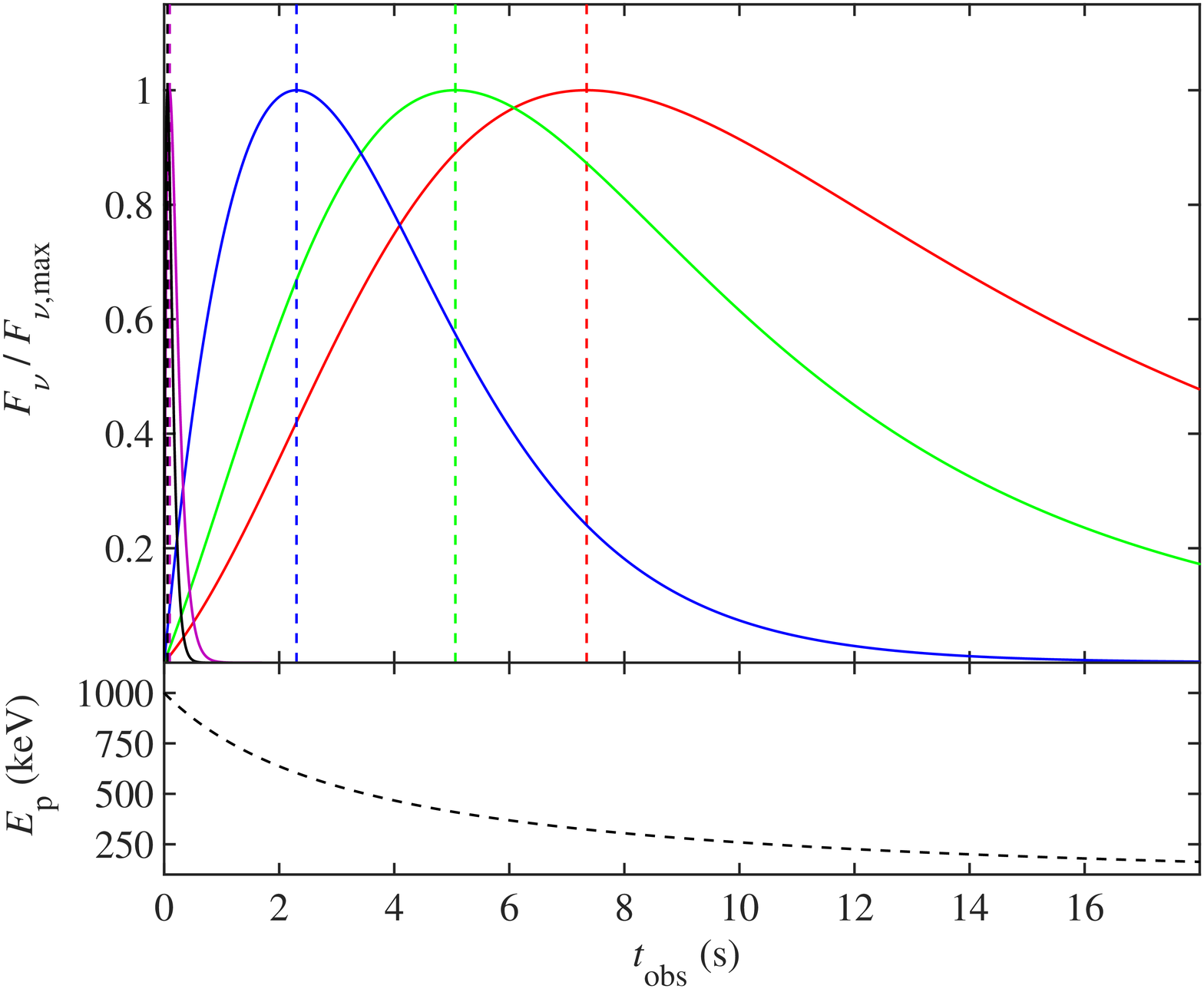} &
\includegraphics[angle=0,scale=0.15]{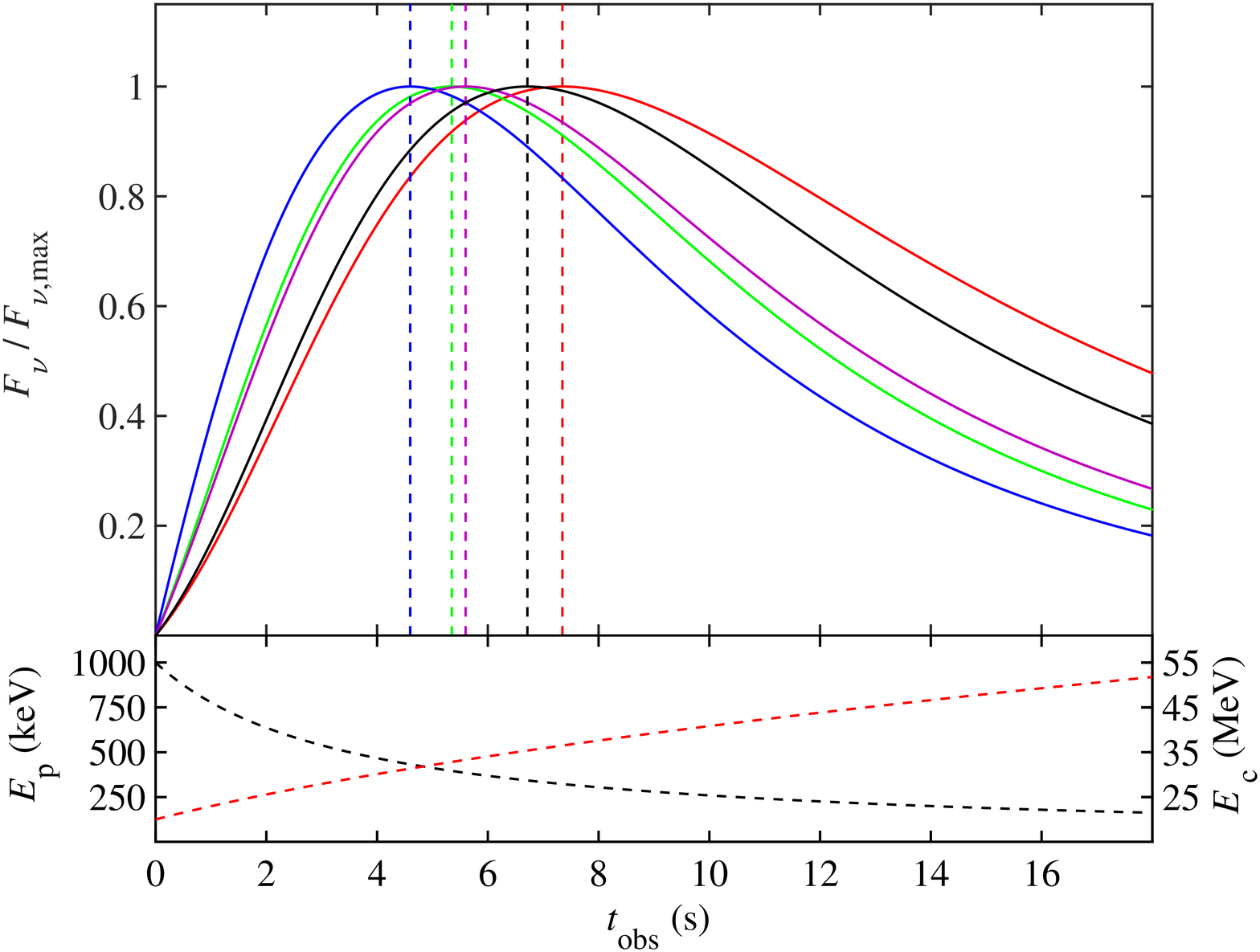} &
\includegraphics[angle=0,scale=0.15]{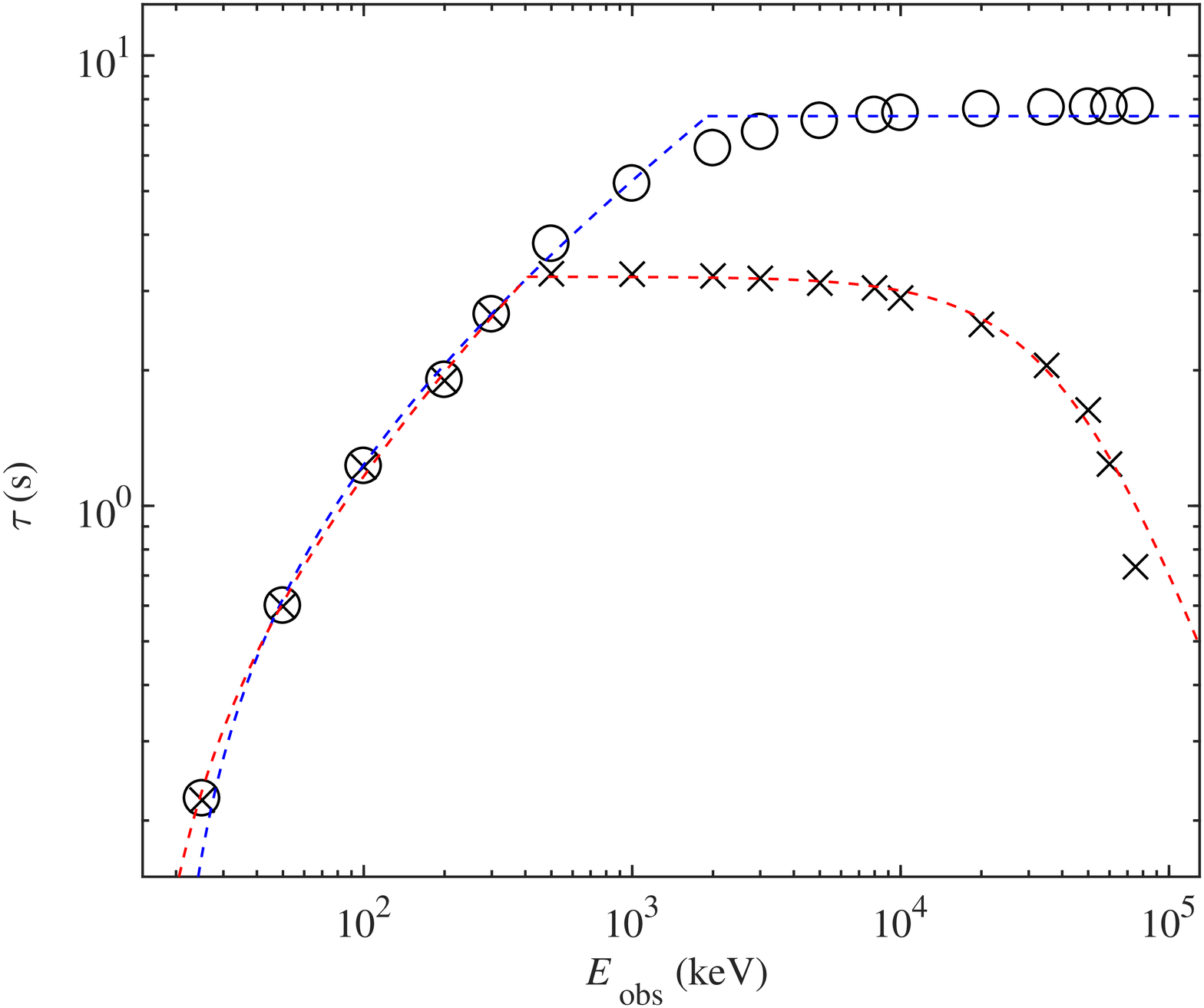} \\
\end{tabular}
\end{center}
\caption{
Left panel: the synthetic light curves (upper sub-figure) and the evolution of $E_{\rm ch}$ (lower sub-figure) for the case with CPL spectrum;
Middle pane: the synthetic light curves (upper sub-figure) and the evolution of $E_{\rm p}$ ($E_{\rm c}$) (lower sub-figure) for the Bandcut case;
Right panel: the $\tau-E_{\rm obs}$ relations with ``$\circ$'' for CPL case and ``$\times$'' for Bandcut case, respectively, and the red dashed lines are fitted by using Equation~(\ref{Eq:tau_Formula}).}
\label{MyFigC}
\end{figure}
%%%%%%%%%%%%%%%%%%%%%%%%%%%%%%%%%%%%%%%0%%%%0%%%%%%%%%0%%%%%%%%%%%%%%%%%%%%%%%%%%%%%%
%%%%%%%%%%%%%%%%%%%%%%%%%%%%%%%%%%%%%%%%%%0%%0%%%%%%0%%%%%%%%%%%%%%%%%%%%%%%%%%%%%%%%%

\clearpage
\begin{figure}
\begin{center}
\begin{tabular}{cc}
\includegraphics[angle=0,scale=0.28]{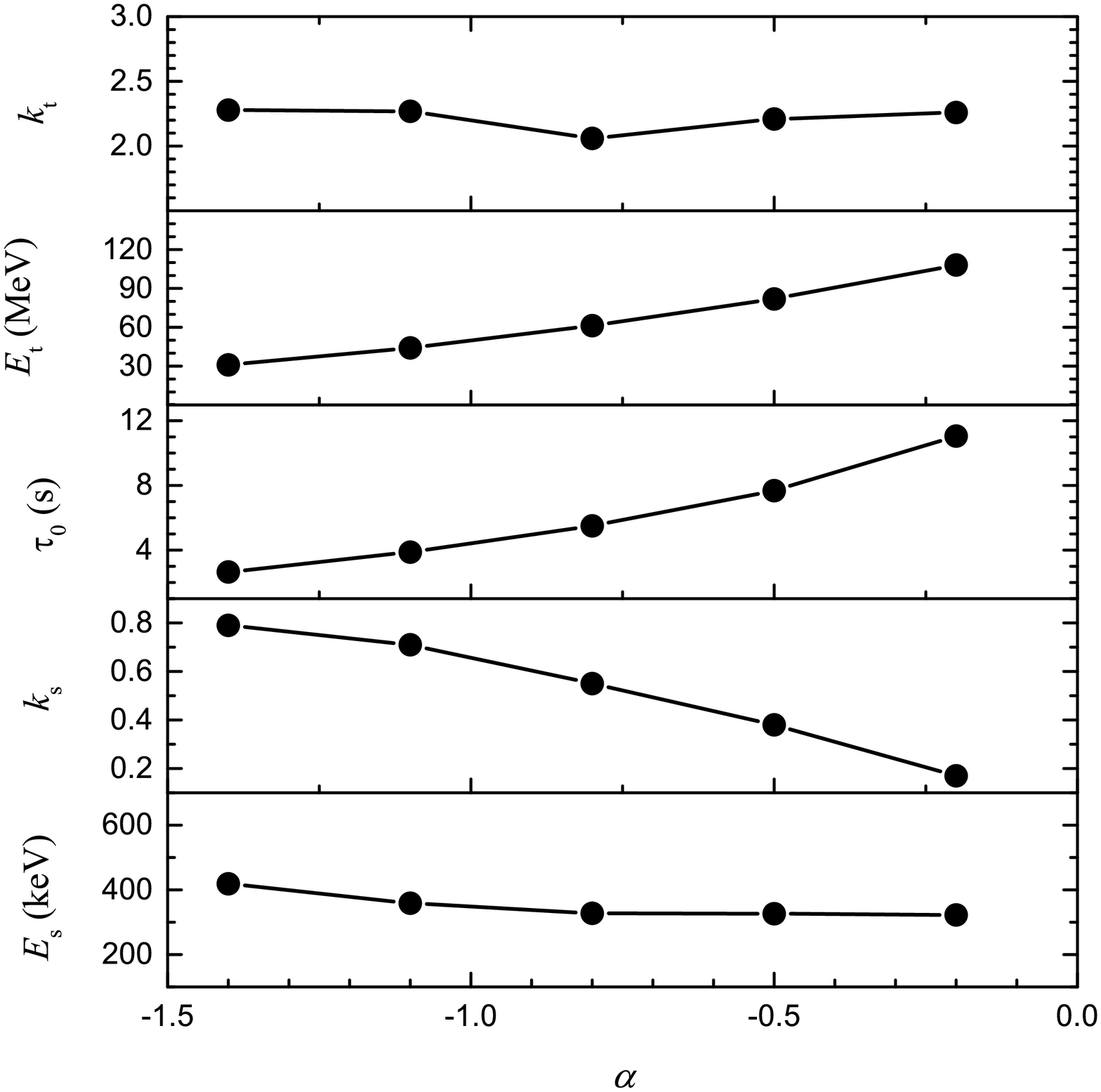} &
\includegraphics[angle=0,scale=0.28]{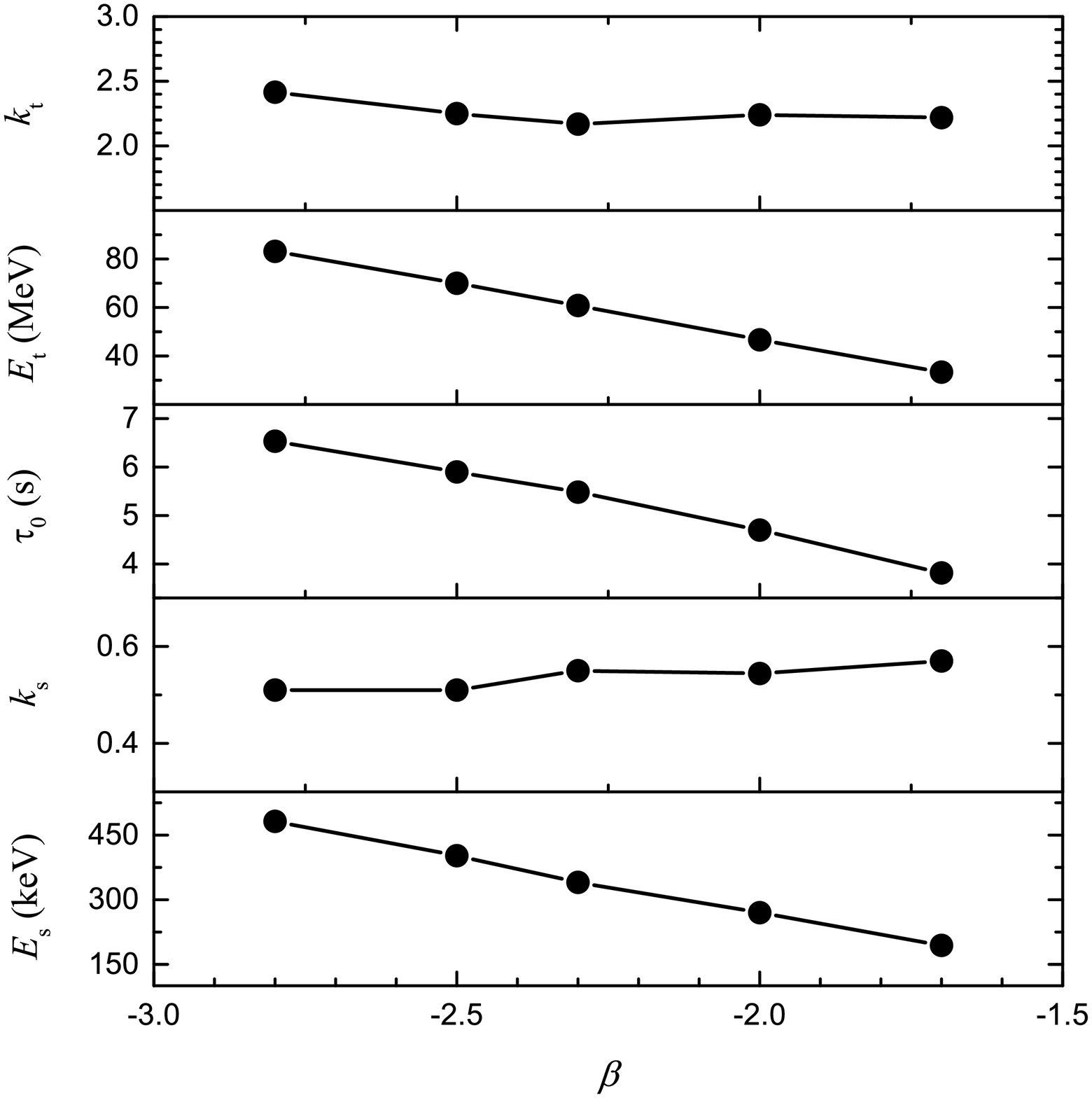} \\
\includegraphics[angle=0,scale=0.28]{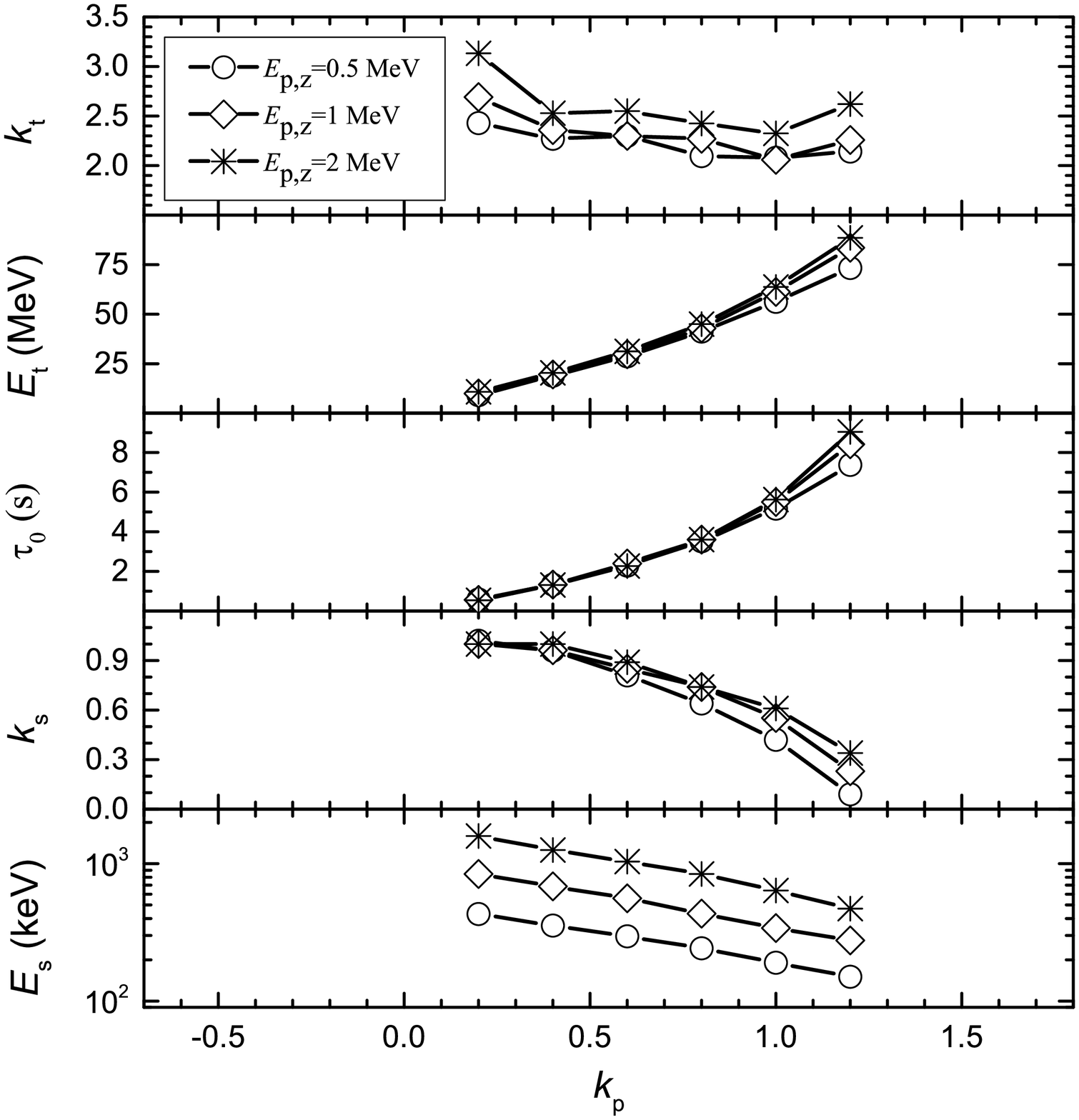} &
\includegraphics[angle=0,scale=0.28]{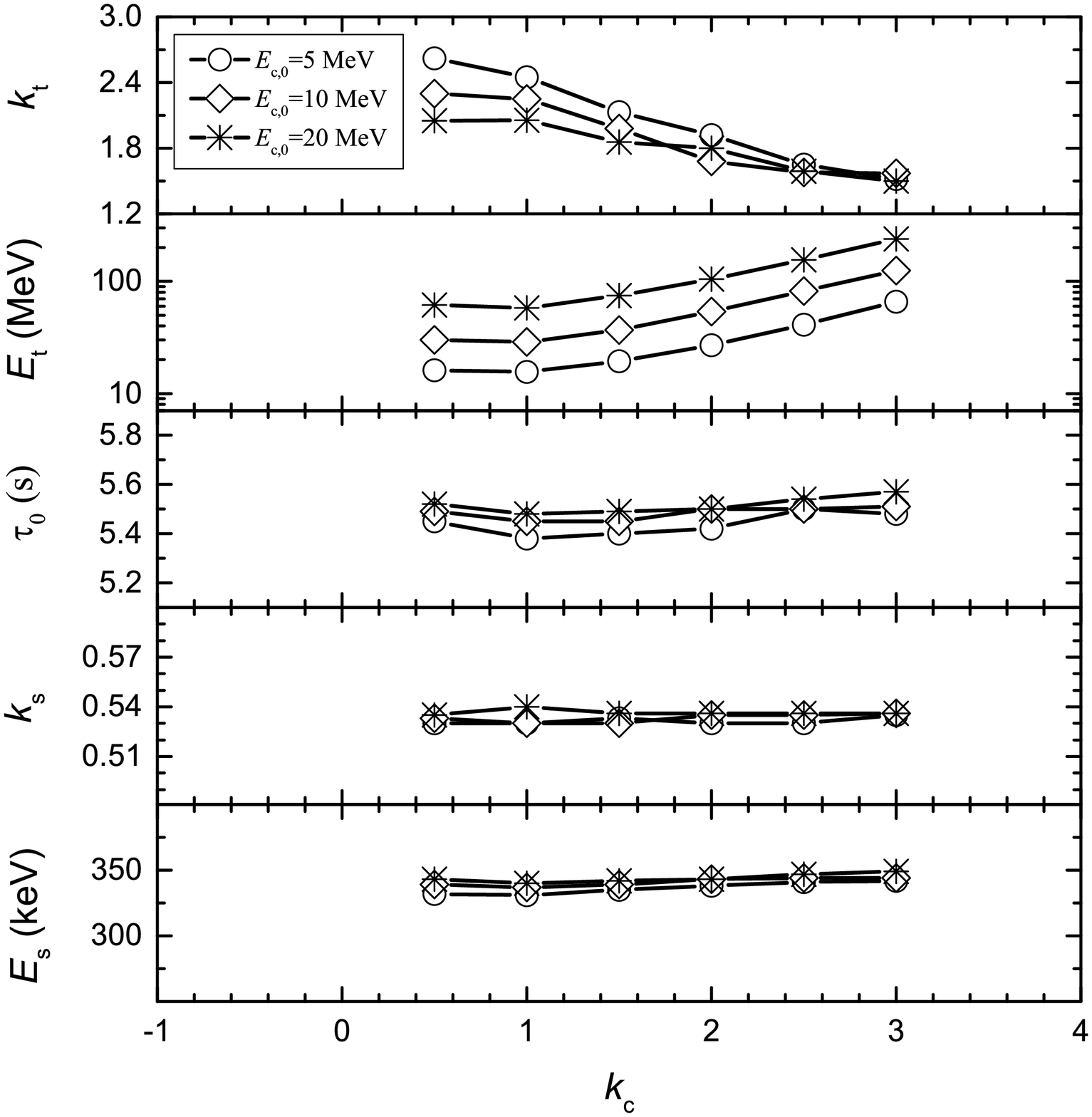} \\
\includegraphics[angle=0,scale=0.302, trim=0 50 0 200]{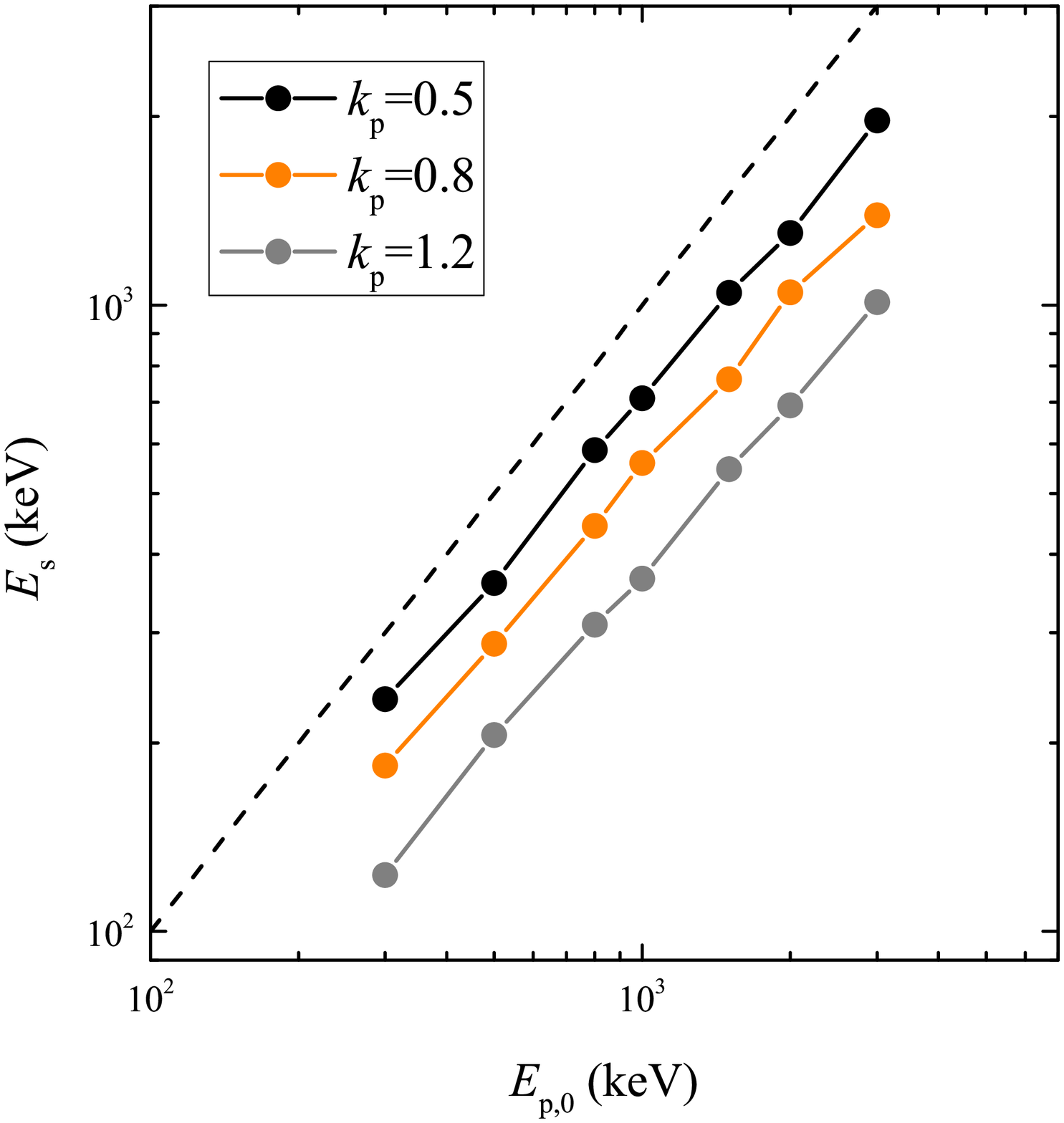} &
\includegraphics[angle=0,scale=0.302, trim=0 50 0 20]{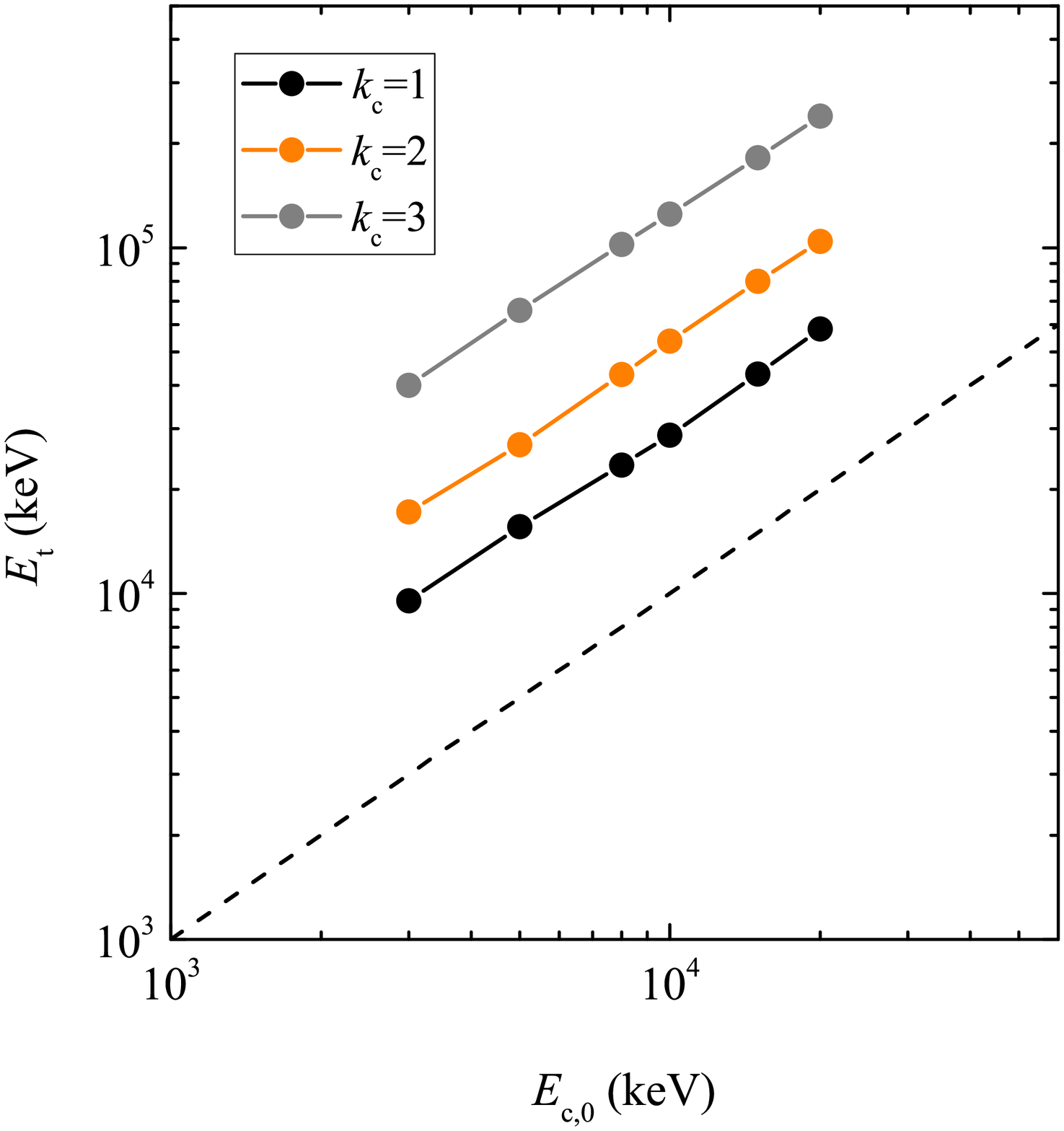} \\
\end{tabular}
\end{center}
\caption{Upper panel: correlations between the spectral lag behavior and $\alpha$ (left) and/or $\beta$ (right) in the Bandcut spectrum;
Middle panel: the dependence of the spectral lag behavior on the $k_{\rm p}$ for different $E_{\rm p,0}$ (left) and $k_{\rm c}$ for different $E_{\rm c,0}$ (right) ;
Lower panel: the relations of $E_{\rm s}-E_{\rm p,0}$ (left) for different $k_{\rm p}$ and $E_{\rm t}-E_{\rm c,0}$ (right) for different $k_{\rm c}$, and the dashed line in each panel denotes the case of $E_{\rm s}=E_{\rm p,0}$ or $E_{\rm t}=E_{\rm c,0}$.}
\label{MyFigD}
\end{figure}
%%%%%%%%%%%%%%%%%%%%%%%%%%%%%%%%%%%%%%%0%%%%0%%%%%%%%%0%%%%%%%%%%%%%%%%%%%%%%%%%%%%%%
%%%%%%%%%%%%%%%%%%0%%%%%%%%%%%%%%%%%%%%%%%%%%0%%%%%%0%%%%%%%%%%%%%%%%%%%%%%%%%%%%%%%%%

\clearpage
\begin{figure}
\begin{center}
\begin{tabular}{ccc}
\includegraphics[angle=0,scale=0.25]{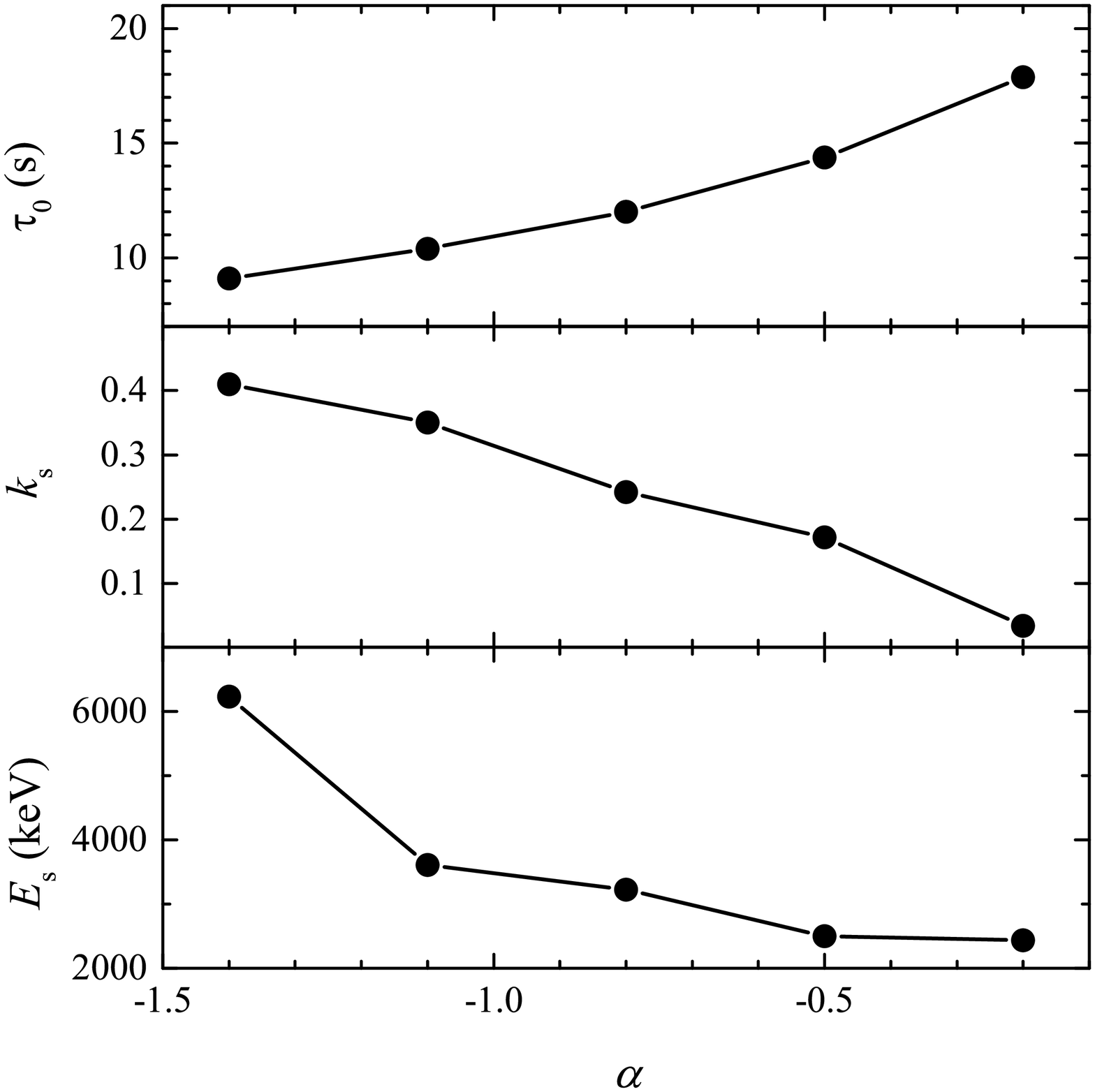} &
\includegraphics[angle=0,scale=0.25,trim=50 0 0 0]{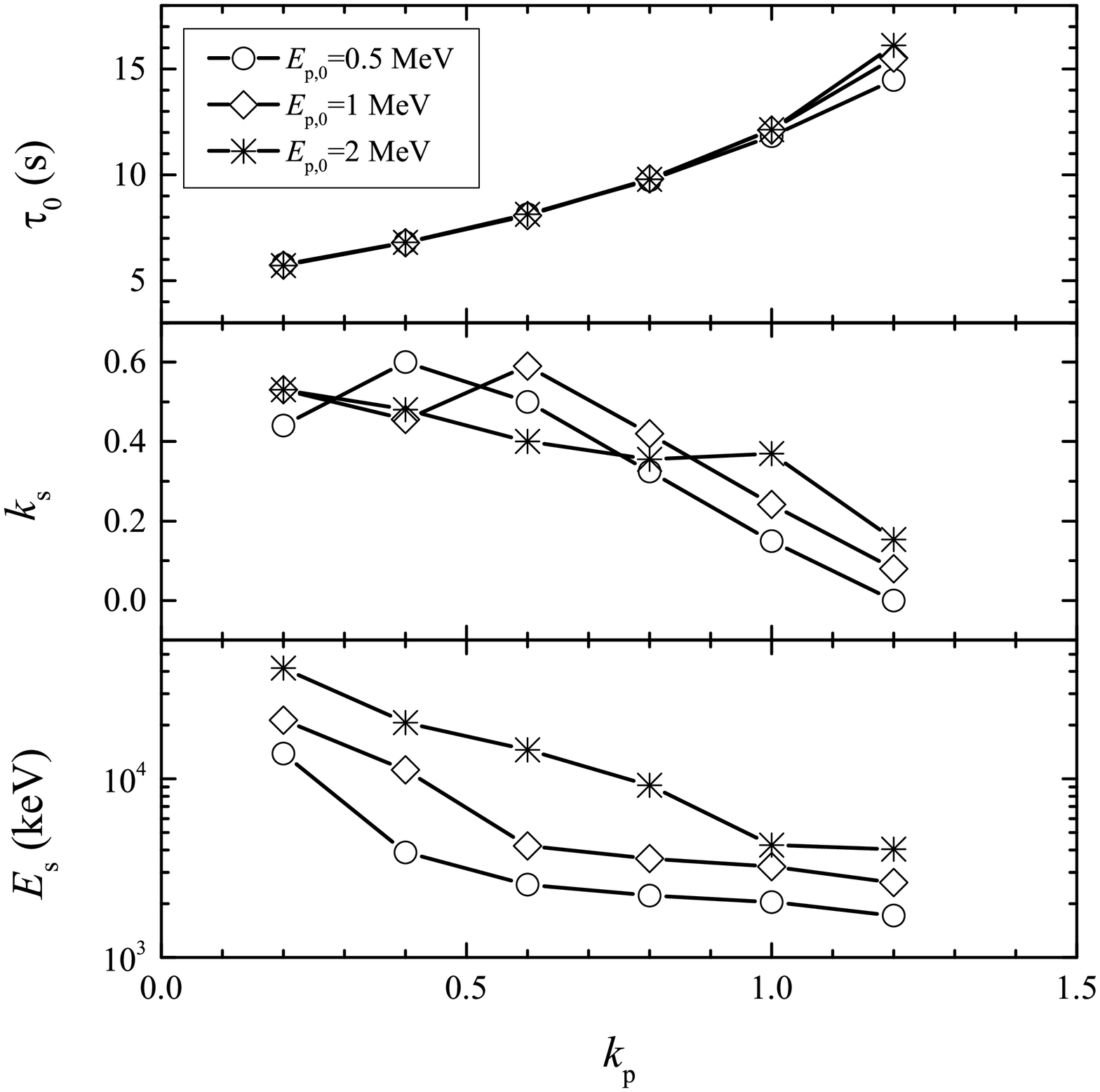} &
\includegraphics[angle=0,scale=0.27,trim=100 36 0 0]{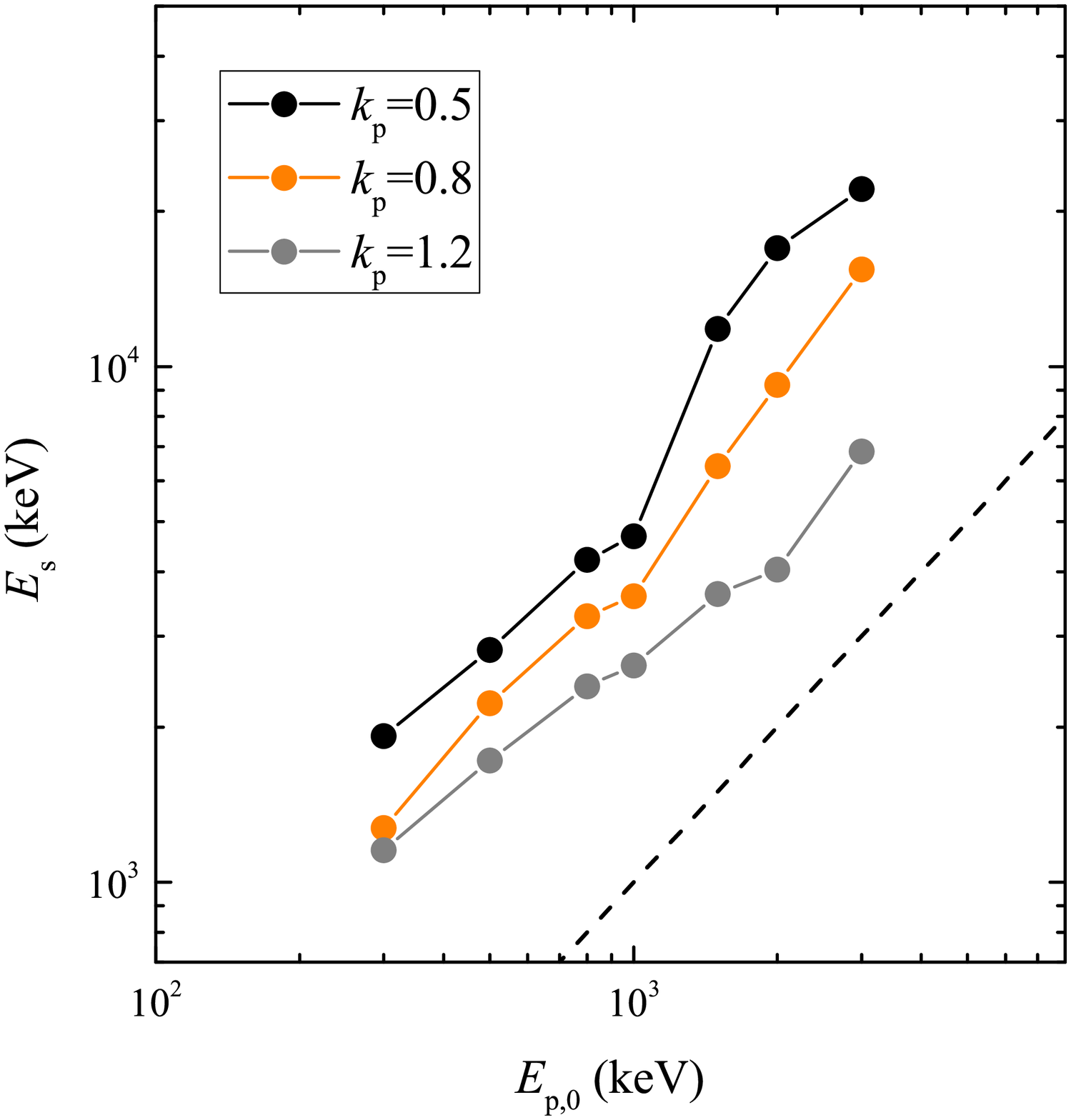}\\
\end{tabular}
\end{center}
\caption{Upper panel: correlations between the spectral lag behavior and $\lambda$ (left) and/or $k_{\rm p}$ (right)  for the case with CPL;
Lower panel: the relations of $E_{\rm s}-E_{\rm p,0}$ for different $k_{\rm p}$, and the dashed line denotes the case of $E_{\rm s}=E_{\rm p,0}$.}
\label{MyFigE}
\end{figure}
%%%%%%%%%%%%%%%%%%%%%%%%%%%%%%%%%%%%%%%0%%%%0%%%%%%%%%0%%%%%%%%%%%%%%%%%%%%%%%%%%%%%%
%%%%%%%%%%%%%%%%%%%%%%%%%%%%%%%%%%%%%%%%%%%%00%%%%%%0%%%%%%%%%%%%%%%%%%%%%%%%%%%%%%%%

\clearpage
\begin{figure}
\begin{center}
\begin{tabular}{cc}
\includegraphics[angle=0,scale=0.315,trim=0 20 0 0]{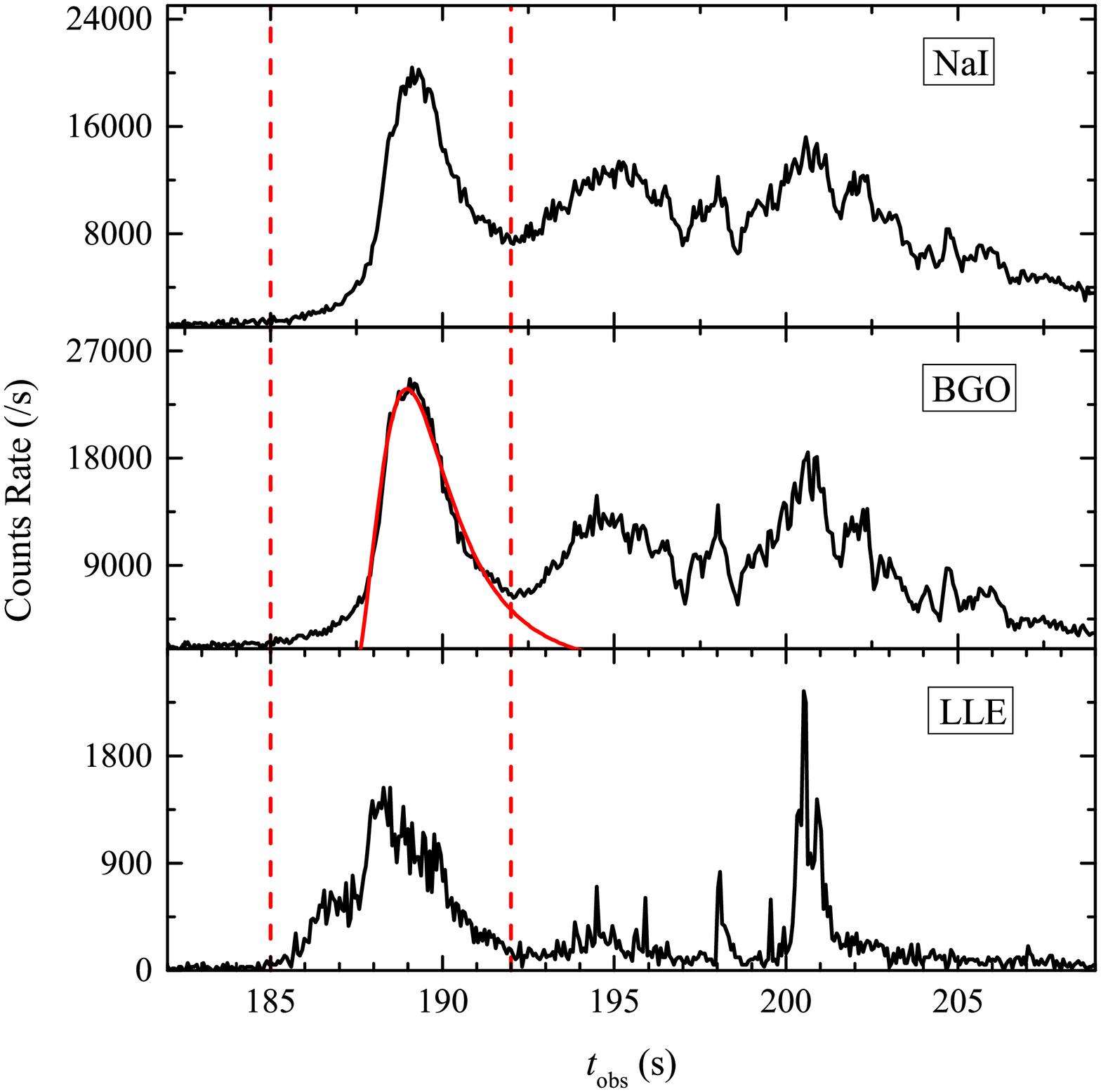} &
\includegraphics[angle=0,scale=0.305]{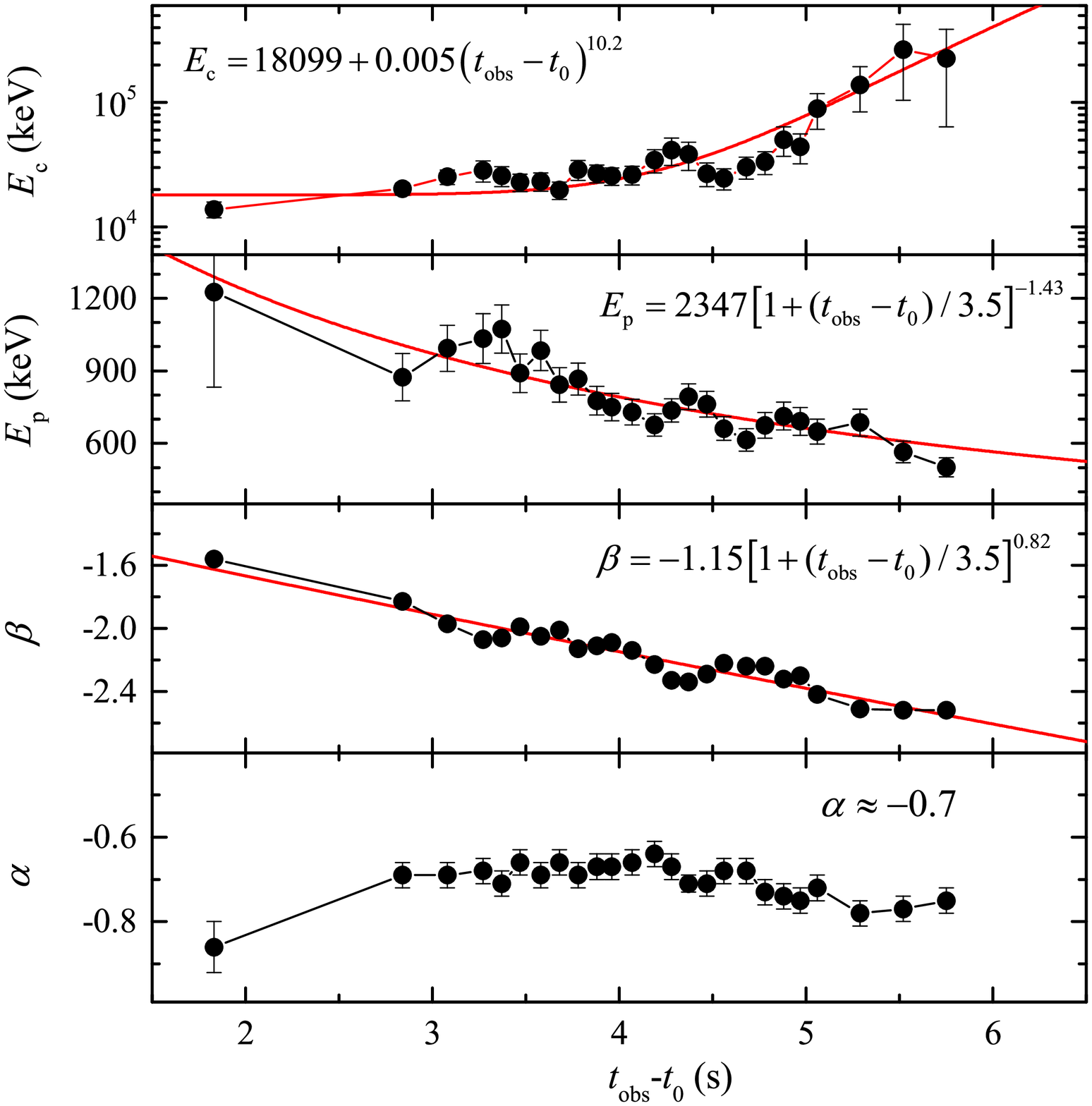} \\
\end{tabular}
\end{center}
\caption{Left Panel: Light curve of GRB~160625B second sub-burst.
Here, the two vertical dashed lines mark the time period for performing the cross-correlation function analysis, and the red curve in middle sub-figure is fitted by using Equation~(\ref{Eq:LC}) and read as  $I_{\rm p,f}=23804$~photons/s, $t_0=187.5$~s, $t_{\rm p,f}=189$~s, $\mathcal{R}=1.27$, and $\mathcal{D}=3.36$.
Right Panel: The temporal evolution of $\alpha$, $\beta$, $E_{\rm p}$, and $E_{\rm c}$.
The red solid lines in each sub-figure denote the fitting results with $t_0=185$~s.
}\label{MyFigF}
\end{figure}
%%%%%%%%%%%%%%%%%%%0%%%%%%%%%%%%%%%0%%%%%0%%0%%0%%%%%%%%%0%%%%%%%%%%%%%%%%%%%%%%%%%%%%%%
%%%%%%%%%%%%%%%%%%0%%%%%%%%%%%%%%0%%%%0%0%0%%%%%%00%%%%%%00%%0%%%%%%%%%%%%%%%%%%%%%%%%%%%%%%%

\clearpage
\begin{figure}
\begin{center}
\begin{tabular}{c}
\includegraphics[angle=0,scale=0.3]{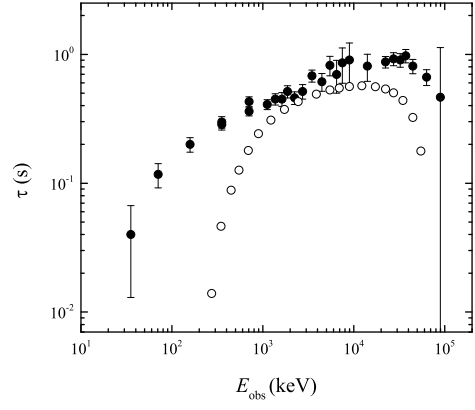}
\end{tabular}
\end{center}
\caption{Right panel: $\tau-E_{\rm obs}$ relation for the first pulse in the GRB~160625B second sub-burst (black dots) and the numerical results of the spectral lag behavior in GRB~160625B by using the empirical model in Section~\ref{Sec:physical origin} is denoted by the `$\circ$' symbol.}
\label{MyFigG}
\end{figure}
%%%%%%%%%%%%%%%%%%%%%%%%%%%%%%%%%%%%%%0%%%%0%%%%%%%%%0%%%%%%%%%%%%%%%%%%%%%%%%%%%%%%
%%%%%%%%%%%%%%%%%0%%%%%%%%%%%%%%%%%%%%%%%%%%0%%%%%%0%%%%%%%%%%%%%%%%%%%%%%%%%%%%%%%%%

\clearpage
\begin{table}
\centering \caption{The time lags between the lowest energy band (10--25~keV)
and any other high energy bands for the second sub-burst of GRB~160625B.}\label{MyTabA}
\begin{tabular}{cc|cc}
\hline
\hline
 Energy & $\tau$ &  Energy & $\tau$ \\
 (keV) & (s) & (keV) & (s)  \\
\hline
25--50	&	$	0.04	\pm	0.027	$	&	5000--6000	&	$	0.823\pm	0.141	$	\\
50--100	&	$	0.117	\pm	0.025	$	&	6000--7000	&	$	0.698	\pm	0.199	$	\\
100--250	&	$	0.2	\pm	0.026$	&	7000--8000	&	$	0.858	\pm	0.262	$	\\
250--500	&	$	0.298	\pm	0.03	$	&	8000--10000	&	$	0.905	\pm	0.32	$	\\
500--1000	&	$	0.43\pm	0.038	$	&	10000--20000	&	$	0.81\pm	0.193	$	\\
1000--1250&	$	0.409	\pm	0.036	$	&	20000--25000	&	$	0.873	\pm	0.089$	\\
1250--1500&	$	0.45	\pm	0.045	$	&	25000--30000	&	$	0.923	\pm	0.106	$	\\
1500--1750&	$	0.45	\pm	0.055	$	&	30000--35000	&	$	0.904\pm	0.11	$	\\
1750--2000&	$	0.515\pm	0.056	$	&	35000--40000	&	$	0.976	\pm	0.116	$	\\
2000--2500&	$	0.46\pm	0.052	$	&	40000--50000	&	$	0.811	\pm	0.104	$	\\
2500--3000&	$	0.516	\pm	0.067	$	&	50000--80000	&	$	0.666	\pm	0.094	$	\\
3000--4000&	$	0.681\pm	0.073	$	&	80000--100000	&	$	0.465	\pm	0.664	$	\\
4000--5000&	$	0.614\pm	0.097	$\\
\hline
\end{tabular}
\label{MyTabA}
\end{table}

%%%%%%%%%%%%%%%%%%%%%%%%%%%%%%%%%%%%%%0%%%%0%%%%%%%%%0%%%%%%%%%%%%%%%%%%%%%%%%%%%%%%
%%%%%%%%%%%%%%%%%%%%%%%%%%%%%%%%%%%%%%%%%%%0%%%%%%0%%%%%%%%%%%%%%%%%%%%%%%%%%%%%%%%%
\clearpage
\bibliography{}

\begin{thebibliography}{}
\expandafter\ifx\csname natexlab\endcsname\relax\def\natexlab#1{#1}\fi

\bibitem[Ackermann et al.(2011)]{Ackermann_M-2011-Ajello_M}Ackermann, M., Ajello, M., Asano, K., et al.\ 2011, \apj, 729, 114
     % Detection of a Spectral Break in the Extra Hard Component of GRB 090926A
     % http://adsabs.harvard.edu/abs/2011ApJ...729..114A
\bibitem[Ackermann et al.(2013)]{Ackermann_M-2013-Ajello_M}Ackermann, M., Ajello, M., Asano, K., et al.\ 2013, \apjs, 209, 11
     % The First Fermi-LAT Gamma-Ray Burst Catalog
     % http://adsabs.harvard.edu/abs/2013ApJS..209...11A

\bibitem[Band et al.(1993)]{Band93} Band, D., Matteson, J., Ford, L., et al.\ 1993, \apj, 413, 281

\bibitem[Band(1997)]{Band97} Band, D.~L.\ 1997, \apj, 486, 928

\bibitem[Baring \& Harding(1997)]{Baring_MG-1997-Harding_AK}Baring, M.~G., \& Harding, A.~K.\ 1997, \apjl, 481, L85
     % Pair Production Absorption Troughs in Gamma-Ray Burst Spectra: A Potential Distance Discriminator
     % http://adsabs.harvard.edu/abs/1997ApJ...481L..85B
\bibitem[Bo{\v s}njak \& Daigne(2014)]{Bosnjak14} Bo{\v s}njak, {\v Z}., \& Daigne, F.\ 2014, \aap, 568, A45

\bibitem[Chang et al.(2008)]{Chang08} Chang, P., Spitkovsky, A., \& Arons, J.\ 2008, \apj, 674, 378

\bibitem[Chen et al.(2005)]{Chen05} Chen, L., Lou, Y.-Q., Wu, M., et al.\ 2005, \apj, 619, 983

\bibitem[Cheng et al.(1995)]{Cheng95} Cheng, L.~X., Ma, Y.~Q., Cheng, K.~S.,
Lu, T., \&amp; Zhou, Y.~Y.\ 1995, \aap, 300, 746

\bibitem[Daigne \& Mochkovitch(2003)]{Daigne03} Daigne, F., \& Mochkovitch, R.\ 2003, \mnras, 342, 587
\bibitem[Deng et al.(2016)]{Deng_Wei-2016-Zhang_Haocheng-ApJ.821L.12D} Deng, W., Zhang, H., Zhang, B., et al.\ 2016, \apj, 821, L12
\bibitem[Fenimore et al.(1993)]{Fenimore_EE-1993-Epstein_RI}Fenimore, E.~E., Epstein, R.~I., \& Ho, C.\ 1993, \aaps, 97, 59
     % The escape of 100 MeV photons from cosmological gamma-ray bursts
     % http://adsabs.harvard.edu/abs/1993A&AS...97...59F
\bibitem[Fishman \& Meegan(1995)]{Fishman_GJ-1995-Meegan_CA-ARA&A.33.415F} Fishman, G.~J., \& Meegan, C.~A.\ 1995, \araa, 33, 415

\bibitem[Fraija(2015)]{Fraija_N-2015}Fraija, N.\ 2015, \apj, 804, 105
     % GRB 110731A: Early Afterglow in Stellar Wind Powered By a Magnetized Outflow
     % http://adsabs.harvard.edu/abs/2015ApJ...804..105F
\bibitem[Fraija et al.(2016)]{Fraija_N-2016-Lee_WH}Fraija, N., Lee, W.~H., Veres, P., \& Barniol Duran, R.\ 2016, \apj, 831, 22
     % Modeling the Early Afterglow in the Short and Hard GRB 090510
     % http://adsabs.harvard.edu/abs/2016ApJ...831...22F
\bibitem[Gao et al.(2012)]{Gao12} Gao, H., Liang, N., \& Zhu, Z.-H.\ 2012, International Journal of Modern Physics D, 21, 1250016-1-1250016-16

\bibitem[Goodman(1986)]{Goodman86} Goodman, J.\ 1986, \apjl, 308, L47

\bibitem[Gehrels et al.(2006)]{Gehrels06} Gehrels, N., Norris, J.~P., Barthelmy, S.~D., et al.\ 2006, \nat, 444, 1044

\bibitem[Ioka \& Nakamura(2001)]{Ioka01} Ioka, K., \& Nakamura, T.\ 2001, \apjl, 554, L163

\bibitem[Keshet et al.(2009)]{Keshet09} Keshet, U., Katz, B., Spitkovsky, A., \& Waxman, E.\ 2009, \apjl, 693, L127
\bibitem[Kobayashi et al.(2007)]{Kobayashi_S-2007-Zhang_B}Kobayashi, S., Zhang, B., M{\'e}sz{\'a}ros, P., \& Burrows, D.\ 2007, \apj, 655, 391
     % Inverse Compton X-Ray Flare from Gamma-Ray Burst Reverse Shock
     % http://adsabs.harvard.edu/abs/2007ApJ...655..391K
\bibitem[Kocevski et al.(2003)]{Kocevski03} Kocevski, D., Ryde, F., \& Liang, E.\ 2003, \apj, 596, 389

\bibitem[Kouveliotou et al.(1993)]{Kouveliotou93} Kouveliotou, C., Meegan, C.~A., Fishman, G.~J., et al.\ 1993, \apjl, 413, L101

\bibitem[Krolik \& Pier(1991)]{Krolik_JH-1991-Pier_EA}Krolik, J.~H., \& Pier, E.~A.\ 1991, \apj, 373, 277
     % Relativistic motion in gamma-ray bursts
     % http://adsabs.harvard.edu/abs/1991ApJ...373..277K
 \bibitem[Lemoine et al.(2013)]{Lemoine13} Lemoine, M., Li, Z., \& Wang, X.-Y.\ 2013, \mnras, 435, 3009

\bibitem[Liang et al.(2006)]{Liang06} Liang, E.~W., Zhang, B., O'Brien, P.~T., et al.\ 2006, \apj, 646, 351

\bibitem[Lin et al.(2018)]{Lin2018} Lin, D.-B., Lu, R.-J., Du, S.-S., et al.\ 2018, \apj, submitted

\bibitem[Lu et al.(2006)]{Lu06} Lu, R.-J., Qin, Y.-P., Zhang, Z.-B., \& Yi, T.-F.\ 2006, \mnras, 367, 275

\bibitem[Lu et al.(2018)]{Lu18} Lu, R.-J., Liang, Y.-F., Lin, D.-B., et al.\ 2018, \apj, 865, 153

\bibitem[McBreen et al.(2008)]{McBreen08} McBreen, S., Foley, S., Watson, D., et al.\ 2008, \apjl, 677, L85

\bibitem[Medvedev et al.(2005)]{Medvedev05} Medvedev, M.~V., Fiore, M., Fonseca, R.~A., et al.\ 2005, \apj, 618, L75

\bibitem[M{\'e}sz{\'a}ros, \& Rees(2000)]{Meszaros_P-2000-Rees_MJ_-ApJ.530.292M} M{\'e}sz{\'a}ros, P., \& Rees, M.~J.\ 2000, \apj, 530, 292
\bibitem[Norris et al.(1986)]{Norris86} Norris, J.~P., Share, G.~H., Messina,
D.~C., et al.\ 1986, \apj, 301, 213

\bibitem[Norris et al.(2000)]{Norris00} Norris, J.~P., Marani, G.~F., \& Bonnell, J.~T.\ 2000, \apj, 534, 248

\bibitem[Norris et al.(2001a)]{Norris01a} Norris, J.~P., Scargle, J.~D., \& Bonnell, J.~T.\ 2001, Gamma-ray Bursts in the Afterglow Era, 40

\bibitem[Norris et al.(2001b)]{Norris01b} Norris, J.~P., Scargle, J.~D., \& Bonnell, J.~T.\ 2001, Gamma 2001: Gamma-Ray Astrophysics, 587, 176

\bibitem[Norris(2004)]{Norris04} Norris, J.~P.\ 2004, Baltic Astronomy, 13, 221

\bibitem[Norris et al.(2005)]{Norris05} Norris, J.~P., Bonnell, J.~T., Kazanas, D., et al.\ 2005, \apj, 627, 324

\bibitem[Norris \& Bonnell(2006)]{Norris06} Norris, J.~P., \& Bonnell, J.~T.\ 2006, \apj, 643, 266

\bibitem[Paczynski(1986)]{Paczynski86} Paczynski, B.\ 1986, \apj, 308, L43

\bibitem[Pe'er(2015)]{Peer_A-2015-AdAst2015E.22P} Pe'er, A.\ 2015, Advances in Astronomy, 2015, 907321

\bibitem[Peng et al.(2007)]{Peng07} Peng, Z.-Y., Lu, R.-J., Qin, Y.-P., \& Zhang, B.-B.\ 2007, \cjaa, 7, 428

\bibitem[Peng et al.(2011)]{Peng11} Peng, Z.~Y., Yin, Y., Bi, X.~W., Bao, Y.~Y., \& Ma, L.\ 2011, Astronomische Nachrichten, 332, 92

\bibitem[Rees \& M{\'e}sz{\'a}ros(1994)]{Rees94} Rees, M.~J., \& Meszaros, P.\ 1994, \apjl, 430, L93

\bibitem[Schaefer(2003)]{Schaefer03} Schaefer, B.~E.\ 2003, \apjl, 583, L67

\bibitem[Schaefer(2007)]{Schaefer07} Schaefer, B.~E.\ 2007, \apj, 660, 16

\bibitem[Shao \& Dai(2005)]{Shao_L-2005-Dai_ZG}Shao, L., \& Dai, Z.~G.\ 2005, \apj, 633, 1027
     % A Reverse-Shock Model for the Early Afterglow of GRB 050525A
     % http://adsabs.harvard.edu/abs/2005ApJ...633.1027S
\bibitem[Shao et al.(2017)]{Shao17} Shao, L., Zhang, B.-B., Wang, F.-R., et al.\ 2017, \apj, 844, 126

\bibitem[Shen et al.(2005)]{Shen05} Shen, R.-F., Song, L.-M., \& Li, Z.\ 2005, \mnras, 362, 59

\bibitem[Shenoy et al.(2013)]{Shenoy13} Shenoy, A., Sonbas, E., Dermer, C., et al.\ 2013, \apj, 778, 3

\bibitem[Silva et al.(2003)]{Silva03} Silva, L.~O., Fonseca, R.~A., Tonge, J.~W., et al.\ 2003, \apj, 596, L121

\bibitem[Tang et al.(2015)]{Tang_QW-2015-Peng_FK}Tang, Q.-W., Peng, F.-K., Wang, X.-Y., \& Tam, P.-H.~T.\ 2015, \apj, 806, 194
     % Measuring the Bulk Lorentz Factors of Gamma-ray Bursts with Fermi
     % http://adsabs.harvard.edu/abs/2015ApJ...806..194T
\bibitem[Thompson(1994)]{Thompson_C_-1994-MNRAS.270.480T} Thompson, C.\ 1994, \mnras, 270, 480
\bibitem[Uhm \& Zhang(2014)]{Uhm14} Uhm, Z.~L., \& Zhang, B.\ 2014, Nature Physics, 10, 351

\bibitem[Uhm \& Zhang(2015)]{Uhm15} Uhm, Z.~L., \& Zhang, B.\ 2015, \apj, 808, 33

\bibitem[Uhm \& Zhang(2016a)]{Uhm16a} Uhm, Z.~L., \& Zhang, B.\ 2016, \apjl, 824, L16

\bibitem[Uhm \& Zhang(2016b)]{Uhm16b} Uhm, Z.~L., \& Zhang, B.\ 2016, \apj, 825, 97

\bibitem[Uhm et al.(2018)]{Uhm18} Uhm, Z.~L., Zhang, B., \& Racusin, J.\ 2018, arXiv:1801.09183

\bibitem[Ukwatta et al.(2010)]{Ukwatta10} Ukwatta, T.~N., Stamatikos, M., Dhuga, K.~S., et al.\ 2010, \apj, 711, 1073

\bibitem[Ukwatta et al.(2012)]{Ukwatta12} Ukwatta, T.~N., Dhuga, K.~S., Stamatikos, M., et al.\ 2012, \mnras, 419, 614

\bibitem[Wei et al.(2017)]{Wei17} Wei, J.-J., Zhang, B.-B., Shao, L., Wu, X.-F., \& M{\'e}sz{\'a}ros, P.\ 2017, \apjl, 834, L13
\bibitem[Woods \& Loeb(1995)]{Woods_E-1995-Loeb_A}Woods, E., \& Loeb, A.\ 1995, \apj, 453, 583
     % Empirical Constraints on Source Properties and Host Galaxies of Cosmological Gamma-Ray Bursts
     % http://adsabs.harvard.edu/abs/1995ApJ...453..583W

\bibitem[Xu et al.(2016)]{Xu16} Xu, D., Malesani, D., Fynbo, J.~P.~U., et al.\ 2016, GRB Coordinates Network, Circular Service, No.~19600, \#1 (2016), 19600, 1

\bibitem[Yi et al.(2006)]{Yi06} Yi, T., Liang, E., Qin, Y., \& Lu, R.\ 2006, \mnras, 367, 1751

\bibitem[Zhang et al.(2009)]{Zhang09} Zhang, B., Zhang, B.-B., Virgili, F.~J., et al.\ 2009, \apj, 703, 1696

\bibitem[Zhang \& Yan (2011)]{Zhang_Yan11} Zhang, B., \& Yan, H. 2011, ApJ, 726, 90

\bibitem[Zhang et al.(2012)]{Zhang12} Zhang, B.-B., Burrows, D.~N., Zhang, B., et
al.\ 2012, \apj, 748, 132
%\bibitem[Zhang et al.(2016)]{Zhang16} Zhang, B.-B., Uhm, Z.~L., Connaughton, V., Briggs, M.~S., \& Zhang, B.\ 2016, \apj, 816, 72
\bibitem[Zhang et al.(2018)]{Zhang18} Zhang, B.-B., Zhang, B., Castro-Tirado, A.~J., et al.\ 2018, Nature Astronomy, 2, 69
\bibitem[Zhang et al.(2018)]{Zhang_BB-2018-Zhang_B-NatCo.9.447Z} Zhang, B.-B., Zhang, B., Sun, H., et al.\ 2018, Nature Communications, 9, 447

\bibitem[Zhao et al.(2014)]{Zhao14} Zhao, X., Li, Z., Liu, X., et al.\ 2014, \apj, 780, 12
\end{thebibliography}

\end{document}